\documentclass[aps,groupaddress,superscriptaddress,reprint,amsmath,amssymb,graphicx]{revtex4-1}
\usepackage{graphicx}
\usepackage{bm}
\usepackage{cancel}
\usepackage{bbold}
\usepackage{tabularx}
\usepackage{booktabs}
\usepackage{xcolor}
\preprint{APS/123-QED}

\newcommand{\A}{\mathrm{a}}
\newcommand{\V}{{\rm v}}
\newcommand{\eq}{\mathrm{eq}}
\newcommand{\ex}{\mathrm{ex}}
\newcommand{\FF}{\mathcal{F}}

\newcommand{\flux}{\mathrm{flux}}
\newcommand{\hk}{\mathrm{hk}}
\newcommand{\KL}{\mathrm{KL}}
\newcommand{\na}{\mathrm{na}}

\newcommand{\st}{\mathrm{st}}
\newcommand{\tot}{\mathrm{tot}}
\newcommand{\average}[1]{\left\langle{#1}\right\rangle}

\newcommand{\D}{\mathrm{d}}
\newcommand{\E}{\mathrm{e}}

\newcommand{\sD}{\mathsf{D}}

\renewcommand{\vec}[1]{\bm{#1}} 

\newcommand{\normD}[2][]{\ifthenelse{\equal{#1}{}}{\left\|{#2}\right\|}{\left\|{#2}\right\|_{#1}}}

\begin{document}

\title{Excess dissipation shapes symmetry breaking in non-equilibrium currents}
\author{Matteo Sireci}
\email{matteo.sireci@phys.ens.fr}
\affiliation{Departamento de Electromagnetismo y F{\'\i}sica de la Materia and Instituto Carlos I de F{\'\i}sica Te\'orica y Computacional, Universidad de Granada, E\text{-}18071 Granada, Spain}
\affiliation{Laboratoire de Physique de l’École Normale Supérieure, CNRS, Paris, France}
\author{Luca Peliti}
\email{luca@peliti.org}
\affiliation{Santa Marinella Research Institute, 00058 Santa Marinella, Italy}
\author{Daniel Maria Busiello}
\email{danielmaria.busiello@unipd.it}
\affiliation{Max Planck Institute for the Physics of Complex Systems, 01187 Dresden, Germany}
\affiliation{Department of Physics and Astronomy “G. Galilei”, University of Padova, Padova, Italy}

\begin{abstract}
\noindent
Most natural thermodynamic systems operate far from equilibrium, developing persistent currents and organizing into non-equilibrium stationary states (NESSs). However, the principles determining how such systems self-organize by breaking equilibrium symmetries to cope with external and internal constraints remain unclear. Here, we establish a general connection between symmetry breaking and dissipation in mesoscopic stochastic systems described by Langevin dynamics. By introducing a geometric framework based on the inverse diffusion matrix, we decompose the velocity field into excess (gradient) and housekeeping (residual) components. This decomposition provides a natural splitting of entropy production, where the excess part measures how the system reorganizes its internal state-space under non-equilibrium conditions, while the housekeeping part quantifies the degree of detailed-balance violation due to external constraints. We derive an exact equality connecting these two parts, and an associated inequality identifying the accessible thermodynamics. We then perform a weak-noise expansion of the stationary solution to systematically characterize the general geometry of the NESS velocity field, enabling a unified classification of non-equilibrium steady states. We apply this framework to a range of systems, from molecular machines to coupled oscillators, showing how symmetry breaking in trajectory space constrains NESS organization. We also extend our approach to systems with multiplicative fluctuations, a challenging and almost uncharted scenario, obtaining a connection between the breaking of additional equilibrium symmetries and the curved (space-dependent) metric. Finally, we show that both the NESS velocity field and stationary distribution can be obtained through variational principles based on excess dissipation.
This work sheds light on the intimate connection between geometric features, dissipative properties, and symmetry breaking, uncovering a classification of NESSs that reflects how emergent organization reflects physical non-equilibrium conditions. Ultimately, our results suggest that excess entropy maximization can serve as a fundamental organizing principle beyond equilibrium thermodynamics.
\end{abstract}

\maketitle

\section{Introduction}
In the natural world, thermodynamic systems are generally away from equilibrium. This particular state might be caused by environmental conditions, such as the presence of multiple thermal baths \cite{busiello2020coarse}, non-conservative forces \cite{seifert2012stochastic}, and odd diffusion \cite{hargus2021odd}, or by internal features, such as energy consuming many-body interactions \cite{Zhang2019TheEC,lynn, falasco2023macroscopic} and chemical dissipation \cite{rao2016nonequilibrium}. In both cases, currents develop in the system, whether of mass, heat, electricity, or molecules, just to mention some possible observables.
How out-of-equilibrium systems tend to self-organize to cope with the external and internal constraints, eventually reaching a stable non-equilibrium stationary state (NESS), is still not understood under general conditions. In other words, a natural question would be to identify an a priori quantity and criterion to identify the thermodynamic configurations that are more probable in non-equilibrium conditions.
Classical general results along this line, such as Onsager's minimum excess dissipation principle \cite{Onsager1,Onsager2} and the Glansdorff–Prigogine criterion \cite{glansdorff1964general}, have attracted recent attention in various contexts \cite{Bertini2004MinimumDP,maes2015revisiting,ito2022information}, originating controversial statements about the role that dissipation should play in shaping the steady state of biological systems \cite{baiesi2018life}. In particular, two thermodynamic quantities, named non-adiabatic and excess entropy production, have been the focus of many works as natural candidates to build general variational principles valid arbitrarily far from equilibrium \cite{jiu1984stability, hatano2001steady,clausius, van2010three, excess_maes}.
Nevertheless, agreement on a unified and solid physical interpretation of such functionals is still lacking, mainly due to the fact that one needs to know the stationary distribution to derive their explicit form. Yet, a possible solution to this conundrum could arise by studying universal physical properties of non-equilibrium systems. In this respect, the last decades have witnessed the discovery of fluctuation theorems \cite{gallavotti1995dynamical,jarzynski1997nonequilibrium,kurchan1998fluctuation,crooks1999entropy,lebowitz1999gallavotti}, universal relations holding arbitrarily far from equilibrium. The core message of these results is that non-equilibrium systems break time-reversal symmetry while preserving a weaker version of it that quantifies dissipation. It is natural to think that the breaking of time symmetry would induce symmetry breaking in other aspects of the system, hinting at the existence of additional universal features. Historically, symmetry-breaking mechanisms are also considered at the heart of the emergence of self-organization in non-equilibrium systems, following the inspiration of Prigogine's idea of ``dissipative structures'' \cite{prigogine1967symmetry,Anderson1987BrokenSE, solvay}. A simple symmetry-breaking phenomenon emerging as a consequence of non-equilibrium conditions is the preferential direction of heat currents between two different thermal baths \cite{Jona_Lasinio_2010, lecomte}. Nevertheless, from a broader and more fundamental perspective, how dissipation induces symmetry breaking in the trajectory space of a general stochastic system is a fascinating topic still largely unexplored. In this work, we make a decisive step towards the understanding of several of these open questions by building a bridge between symmetry breaking, the physical meaning of excess dissipation, and the onset of self-organization of non-equilibrium systems reaching a NESS.

\subsection{Outline of the work}
We focus on continuous-space mesoscopic systems described by Stochastic Thermodynamics \cite{peliti2021stochastic}, a framework encompassing well-known, experimentally realizable examples from molecular machines \cite{seifert2012stochastic} to chemical reaction networks \cite{rao2016nonequilibrium} and colloidal particles \cite{ciliberto2017experiments}.\\ 

In Sections \ref{sec:geom_dis}–\ref{sec:dyn}, we present a general framework based on Langevin dynamics with additive noise. In particular, in Section \ref{sec:geom_dis}, by adopting the (inverse) diffusion matrix as a natural metric in configuration space, we formulate thermodynamics in terms of the current velocity and a generalized free energy (\emph{availability}), revealing their geometric and thermodynamic roles. This setting naturally decomposes the velocity into excess (gradient) and housekeeping (residual) parts. While the housekeeping component measures the breaking of detailed balance due to external conditions, the excess one reflects the system’s self-organization, connected to the relative entropy with respect to equilibrium. This leads to a decomposition of entropy production, implying that, at stationarity, its value equals the housekeeping dissipation \emph{minus} the excess one. From these results, we derive an upper bound for the steady-state dissipation and interpret the excess contribution as energy absorbed from the environment to reshape the NESS. In Section \ref{sec:dyn}, we use general minimization principles to show how thermodynamics shapes symmetry breaking. We generalize the Glansdorff–Prigogine criterion by studying the \emph{non-adiabatic} excess entropy production, determining the dynamics to reach the steady state.
Complementarily, we define an Onsager-like excess variational functional to determine the velocity geometry, once the NESS is given. We show that both the excess and non-adiabatic functionals  split into a term proportional to total entropy production and another related to directional dissipation which drives symmetry breaking. A path-integral analysis further reveals that the housekeeping contribution quantifies the stress–energy tensor asymmetry and is related to cycling trajectories, while excess dissipation selects occupancy of higher-energy states.
We also obtain a variational principle satisfied by
the NESS, namely as the maximal principle of the excess entropy constrained by the value of the Jarzynski-like statistics of the
housekeeping entropy production.
Overall, these three complementary principles reveal that any NESS can be interpreted as the minimally dissipative state consistent with fluxes imposed by external non-equilibrium conditions, highlighting the minimization of excess or non-adiabatic entropy production as potentially more fundamental than the second law of thermodynamics for non-equilibrium systems \cite{van2010three,hatano2001steady,sekimoto1998langevin,ito2022information,dechant2022geometric,ito_non_linear,maes_minimum}.\\

In Sections \ref{sec:NESS}–\ref{sec:osc}, we apply this framework to classify NESS thermodynamics through their geometric structures. Specifically, in Section \ref{sec:NESS}, we name \emph{locally closed} a system characterized by closed fluxes in the trajectory space. When also excess fluxes are closed, we name the system \emph{locally balanced}. Finally, system in \emph{local equilibrium} satisfy a generalized Maxwell–Boltzmann stationary distributions and the availability minimum coincides with the equilibrium one. In Section \ref{sec:noise_exp}, a small-noise expansion of the adiabatic availability shows that, at lowest order, the adiabatic velocity remains divergence-free, constraining the housekeeping and excess geometry. By increasing the noise strength, the excess terms can develop new thermodynamic fluxes with non-zero divergence through coupling with the heat bath. In Section \ref{sec:multiplicative}, we extend the framework to multiplicative noise, where variable-dependent diffusion induces non-zero curvature in state space, adding a "Thermophoretic-like" term to the excess velocity and stabilizing high-energy states. In Section \ref{sec:underdamped}, we study the underdamped case in a temperature gradient, emphasizing the role of total energy changes. Finally, Section \ref{sec:osc} analyzes many-body coupled molecular oscillators undergoing a synchronization phase transition, revealing that housekeeping dissipation links to control parameters while the excess component arises from the order parameter, shaping a complex velocity geometry reflecting free-energy transduction.

\section{Dissipation and geometry of non-equilibrium currents}\label{sec:geom_dis}
\subsection{Thermodynamics of availability\\and current velocity}\label{sec:geom_diff_av}
In this section, we set the mathematical stage by formulating stochastic thermodynamics in a geometric perspective using the inverse diffusion matrix as a metric \cite{Graham1977CovariantFO}. Then,  we introduce the main actors: the availability (or non-equilibrium free energy) and the current velocity \cite{graham1984weak, wang_gauge, wang_rev}.

We consider a system in $d$ dimensions, whose state is described by the variable $\vec{x}=(x^i)$, which evolves according to the Langevin equation in configuration space:
\begin{equation}\label{eq:Langevin}
\frac{\D \vec{x}}{\D t}=\vec{F}(\vec{x})+\vec{\xi}(t),
\end{equation}
where $\vec{F}$ is an external drift, which may be conservative (i.e., stemming from a potential energy) or non-conservative, and $\vec{\xi}(t)=\left(\xi^i(t)\right)$ is a zero-mean Gaussian white-noise process which satisfies
\begin{equation}\label{eq:diffMatrix}
\average{\xi^i(t)\xi^j(t')}_\xi=2 D^{ij}\,\delta(t-t'),\qquad i,j=1,\ldots, d,
\end{equation}
with a given, symmetric and positive-definite diffusion matrix $\sD
=\left(D^{ij}\right)$. We write the indices as exponents for further convenience and denote by $\average{\ldots}_\xi$ the average taken with respect to the noise. In the present case, the diffusion matrix is independent of $\bm x$, hence the manifold of the configuration space is flat. 
We define the \emph{diffusive} scalar product $\bm{a}*\bm{b}$ between the two vectors $\bm{a}=(a^i)$ and $\bm{b}=(b^i)$ by
\begin{equation}\label{eq:diffusive_scalar_prod}
\vec{a}*\vec{b}=a^i D_{ij} b^j,
\end{equation}
where $D=(D_{ij})$ is the matrix inverse of $\sD=\left(D^{ij}\right)$. Throughout the whole manuscript, we use the Einstein convention for summing over repeated indices, one upper and one lower. Thus, $\sD^{-1} = D$ acts as the metrics in the state space, as already noted in \cite{Graham1977CovariantFO,polettini2013generally}. This geometric structure can be viewed as a generalization of the factor $\beta = 1/k_{\rm B}T$, with the inverse diffusivities entering naturally in the diffusive scalar product, i.e., the effective length-scale along each direction is inversely proportional to its diffusivity. Then, the derivative of any scalar field $\phi(x)$ is straightforwardly defined as  $\partial_i \phi$. Likewise, the \emph{covariant} derivative of a vector field $\vec{w}=(w^i)$ is given by $\partial_j w^i$.
On the other hand, the \emph{contravariant} derivative of the scalar field $\phi$ is given by
\begin{equation}\label{eq:gradient}
\left(\nabla \phi\right)^i=D^{ij}\partial_j \phi,
\end{equation}
which is often called the gradient of $\phi$. Finally, we consider the divergence of a vector field as the covariant derivative summed with respect to repeated indices, i.e., $\nabla*\vec{w}=\partial_i w^i$, which coincides with the scalar product of the gradient and the field. With this formalism, the Langevin equation \eqref{eq:Langevin} corresponds to the following Fokker-Planck equation for the probability $P$ of finding the system at time $t$ in the state $\bm x$:
\begin{equation}
\partial_t P=-\nabla *\left(\vec{F}\,P-\nabla P\right)=-\nabla* \vec{J},
\end{equation}
where the probability current $\vec{J}=\left(J^i\right)$ is given by
\begin{equation}
J^i=F^i\,P - D^{ij} \partial_j P=F^i-\left(\nabla P\right)^i.
\end{equation}
Since we do not consider boundary-driven systems for the time being, we take boundary conditions for which $P\to 0$ for $\left|\bm{x}\right|\to\infty$ or for $\bm{x}$ approaching the boundary of the allowed state space. 

Given a time-independent drift $\vec{F}(\bm{x})$, the system eventually reaches a stationary distribution, $P_{\rm st}$, which can be characterized in terms of an non-equilibrium free energy $\FF(\bm{x})$ (hereafter called \emph{availability} \cite{Callen:450289}): 
\begin{equation}\label{eq:availability}
P_{\st}(\bm{x}) \propto \E^{-\FF(\bm{x})}
\end{equation}
and which satisfies $\partial_t P = 0$. Then, we define the current velocity $\vec{v}(\bm{x}, t)$ via the relation
\begin{equation}
\vec{v}=\frac{\vec{J}}{P},
\end{equation}
that becomes equal to $\vec{v}_{\rm st}(\bm{x})$ at stationarity.
In terms of this steady current velocity, the stationarity condition takes the form
\begin{equation}\label{eq:stationary}
\nabla* \vec{v}_\st = \vec{v}_\st * \nabla\FF=\nabla_{v_\st}\FF.
\end{equation}
The lhs represents the divergence of the stationary velocity, while the rhs represents the derivative of the availability with respect to the local velocity, i.e., the rate of change of $\FF$ along the current.

Following a standard approach in thermodynamics, we evaluate the time derivative of the average of a generic thermodynamic quantity $A(x,t)$ by
\begin{eqnarray}\label{eq:av_change_flux}
   d_t \langle A\rangle=\langle \vec{v}*\nabla A\rangle+\langle\dot{A}\rangle,
\end{eqnarray}
where $\average{\ldots}$ denotes the average over the probability distribution:
\begin{eqnarray}
    \average{\ldots}=\int \D x\; \ldots\, P(x).
\end{eqnarray}
The change induced by the explicit time dependence of $A$ is due to driving mechanisms, while the one along the velocity defines the flux of $A$ and emerges by averaging the change along a single trajectory over the noise \cite{seifert_entropy,busiello2019hyperaccurate}:
\begin{eqnarray}
    A_{\rm flux}=\langle \vec{v}* \nabla A(\bm{x})\rangle=\langle \vec{\dot{x}}*\nabla A(\bm{x})\rangle_{\xi}.
\end{eqnarray}
Note that a positive sign indicate a flux from the system to the environment and viceversa for a negative sign.
As a particular case, if we consider the system Gibbs-Shannon entropy $S_{\rm sys} = - \int \D x\; P\log P$, its time derivative 
decomposes into the total and environmental entropy productions that, in our notation, take the following form \cite{wang_rev}:
\begin{eqnarray}\label{eq:entropy_flux}
\dot{S}_{\rm sys}&=& 
\dot{S}_{\rm tot}-\dot{S}_{\rm env},\\
    \dot{S}_{\tot}&=&\int\D x\;P\,v^iD_{ij} v^j=\average{\vec{v} * \vec{v}}\geq 0,\\
\dot{S}_{\rm env}&=&\int\D x\;P(x)\,v^i D_{ij} F^j=\average{\vec{v} * \vec{F}}.
\end{eqnarray}

\subsection{The geometry of the velocity field}\label{sec:geom}
Eq.~\eqref{eq:entropy_flux} suggests that the velocity plays a more fundamental, yet underappreciated, role than the current $\vec{J}$. In particular, Eq.~\eqref{eq:stationary} implies that, while the stationary current is necessarily divergence-free, the stationary velocity is not.
Therefore, we expect $\vec{v}_{\rm st}$ to exhibit a richer geometric structure with non-trivial interplay with dissipative thermodynamic quantities that characterize a non-equilibrium steady-state (NESS). Specifically, here we show two orthogonality conditions for the velocity field that represent the first steps of this study.
Following
\cite{van2010three}, we start by decomposing the velocity into two contributions, one reflecting the steady non-equilibrium condition (i.e., \emph{adiabatic}, $\vec{v}_{\rm a}$) and the other the time-dependent  drivings or relaxation (i.e., \emph{non-adiabatic}, $\vec{v}_{\rm na}$):
\begin{equation}\label{eq:velocity_adiabatic_dec}
\vec{v}=\vec{v}_{\rm a}+\vec{v}_{\rm na} \;.
\end{equation}
Given the absence of time-dependent driving, these two contributions are given by
\begin{eqnarray}
\label{eq:va}
v_{\rm a}^i&=&v^i_{\st} = F-D^{ij}\partial_j\ln P_{\st}=F+D^{ij}\partial_j\mathcal{F},\\
\label{eq:vna}
v_\na^i&=&-D^{ij}\partial_j\ln \frac{P}{P_\st}=D^{ij}\partial_j\mathcal{F}_{\na}(x,t) \;,
\end{eqnarray}
where we have defined the non-adiabatic availability, $\mathcal{F}_\na$. Thanks to these definitions, we can rewrite the Fokker-Planck equation solely in terms of velocity and time-dependent availability, $\FF(t)=\FF+\FF_{\na}(t)=-\log P(x,t)$:
\begin{eqnarray}\label{eq:fp_div}
\partial_t\FF(t)+\vec{v}*\nabla\FF(t)=\nabla*\vec{v},
\end{eqnarray}
By taking the average of Eq.~\eqref{eq:fp_div}, we are able to connect the change in system entropy to the average velocity divergence:
\begin{eqnarray}\label{eq:entropy_div}
    \dot{S}_{\rm sys}=\D_t\average{\FF(t)}=\average{\vec{v}*\nabla\FF(t)}=\average{\nabla*\vec{v}}.
\end{eqnarray}
Equations \eqref{eq:stationary}, \eqref{eq:fp_div}, and \eqref{eq:entropy_div} show that the divergence of the velocity field carries a thermodynamic meaning: a positive divergence corresponds to entropy/availability-increasing expansions, while a negative divergence corresponds to compressions that store work in the system. 

Using Eqs.~\eqref{eq:va} and \eqref{eq:vna}, directly from the stationarity condition, Eq.~\eqref{eq:stationary}, we obtain that the adiabatic and non-adiabatic velocities are on average orthogonal to each other at any time:
\begin{equation}\label{eq:perp_na}
\average{\vec{v}_{\rm a} * \vec{v}_\na}=\int\D x\;P\,v^i_{\rm a}D_{ij}v^j_\na=0,\qquad \forall t.
\end{equation}
Eq.~\eqref{eq:perp_na} implies the standard decomposition of the entropy production into adiabatic and non-adiabatic contributions as well:
\begin{equation}\label{eq:entropy_adiabatic}
\dot{S}_\tot=\average{\vec{v}*\vec{v}}=\dot{S}_{\rm a}+\dot{S}_{\na},
\end{equation}
where 
\begin{align}
\dot{S}_{\rm a}&=\average{\vec{v}*\vec{v}_{\rm a}}=\average{\vec{v}_{\rm a} * \vec{v}_{\rm a}}\geq 0 \;,\\
\dot{S}_\na&=\average{\vec{v} * \vec{v}_\na}=\average{\vec{v}_\na * \vec{v}_\na}\ge 0 \;.
\end{align}
At the stationary state, the non-adiabatic entropy production vanishes, and thus the adiabatic one yields the total entropy production:
\begin{eqnarray}
    \dot{S}^{\st}_{\na}&=&0 \;, \qquad \dot{S}^{\st}_{\rm tot}=\dot{S}^{\st}_{\rm a} \;.
\end{eqnarray}

To derive the second orthogonality condition, let $\phi(\bm{x})$ be an arbitrary time-independent scalar field.
From Eq.~\eqref{eq:av_change_flux}, we obtain that the derivative of its average is given by its flux:
\begin{equation}
\frac{\D}{\D t}\average{ \phi}=\average{\vec{v}*\nabla\phi}=\phi_{\rm flux}.
\end{equation}
Since its average is constant at stationarity, we obtain
\begin{equation}\label{eq:gradient_perp}
\phi_{\rm flux}^\st = \average{\vec{v}_{\rm a}* \nabla\phi}^\st=0 \;.
\end{equation}
This holds for any $\phi$ and in particular for the special choice $\phi=\FF$. Therefore, Eq.~\eqref{eq:gradient_perp} implies that, at the NESS, the adiabatic velocity is orthogonal on average to the contour lines of the availability:
\begin{eqnarray}\label{eq:adiabatic_availability_perp}
     \frac{\D}{\D t}\average{\FF}^\st=\FF_{\rm flux}^\st=\average{\vec{v}_{\rm a}*\nabla\FF}^\st=0 .
\end{eqnarray}

Finally, from Eqs.~\eqref{eq:va} and \eqref{eq:adiabatic_availability_perp}, the drift $\bm{F}$ can be naturally decomposed into two  (on average) orthogonal parts, one dissipative, $\vec{v}_{\rm a}$, and one conservative, $\nabla \mathcal{F}$:
\begin{eqnarray}\label{eq:force_dec}
    \vec{F}=\vec{v}_{\rm a}-\nabla\FF \;.
\end{eqnarray}

\subsection{House-keeping and excess decomposition\\of the velocity}\label{sec_:ex_hk}
The necessary condition for reaching an equilibrium state is detailed balance, or, equivalently, a vanishing adiabatic current, $\vec{J}_{\A}=0$. This implies, from Eq.~\eqref{eq:force_dec}, that the force stems solely from the gradient of a Boltzmann distribution, $\bm{F}=\nabla P_{\mathrm{eq}}$. In the presence of only one physical current and assuming the validity of Einstein's relation (we generalize later to more complex scenarios), this condition reduces to:
\begin{eqnarray}\label{eq:db1}
        \partial_i F_j= \partial_j F_i\;,
\end{eqnarray}
where $F_{i}=(D^{-1}\vec{F})_{i}$. Eq.~\eqref{eq:db1} specifies that the necessary condition for having an equilibrium state, and hence no adiabatic dissipation, is to derive the force $\vec{F}$ from a scalar potential that can be identified as the equilibrium distribution. The orthogonality condition in Eq.~\eqref{eq:gradient_perp} suggests to introduce a generalized detailed balance condition for NESSs by using Eq.~\eqref{eq:force_dec} and splitting dissipative terms breaking detailed balance from those that can be written as gradient of a scalar potential and, as such, that can be ``reabsorbed'' into the gradient of an effective equilibrium distribution. By doing so, we find a new decomposition for the adiabatic velocity into an \emph{excess} part (indicated by $\vec{v}_{\ex, \rm a}$), containing all the gradient terms 
and a residual part that we name \emph{house-keeping} (and indicate by $\vec{v}_{\hk}$):
\begin{eqnarray}\label{eq:dec_hk_ex_a}    \vec{v}_{\A}=\vec{v}_{\hk}+\vec{v}_{\ex,\A}=\vec{v}_{\hk}+\nabla\FF_{\ex,\A} \;.
\end{eqnarray}
The scalar field contributing to the excess adiabatic contribution is named $\mathcal{F}_{\rm ex, a}$. This excess term still satisfies a generalized detailed balance condition as in Eq.~\eqref{eq:db1}:
\begin{eqnarray}\label{eq:gen_db1}
v^{i}_{\ex,\rm a}&=&D^{ij}\partial_{j}\FF_{\ex,a}\;,\\
\label{eq:gen_db2}
\partial_iD_{jk} v^{k}_{\ex,\rm a}&=&  \partial_jD_{ik} v^{k}_{\ex,\rm a} \;.
\end{eqnarray}
We remark that this decomposition is in general terms not unique. Yet, 
as we will see in the next section, physical motivations justify the correct choice, analogously to a gauge fixing,
making $\mathcal{F}_{\rm ex, a}$ the difference in availability induced by non-equilibrium conditions at steady state.
Since the adiabatic velocity only accounts for the stationary contributions, when complemented with the time-dependent part, this new decomposition gives:
\begin{eqnarray}\label{eq:excess_time_dec}
\vec{v}&=&\underbrace{\vec{v}_{\hk}+\vec{v}_{\ex,\A}}_{\vec{v}_\A} \, + \, \vec{v}_{\na} \equiv \vec{v}_{\hk}+\vec{v}_{\ex}
\end{eqnarray}
where we have defined
\begin{eqnarray}\label{eq:def_excess}
\vec{v}_{\ex}&=&\vec{v}_{\ex,\A}+\vec{v}_{\na}=\nabla \FF_{\ex} \;.
\end{eqnarray}
in which the time-dependent excess availability $\FF_\ex(\bm{x},t)$ is defined by
\begin{eqnarray}\label{eq:ex_na}
    \FF_{\ex}=\FF_{\ex,\A}+\FF_{\na} \;.
\end{eqnarray}
where the nonadiabatic availability is defined by Eq.~\eqref{eq:vna}.

\subsection{Housekeeping and excess dissipation encode external conditions and internal organization}
To shed light on the physical meaning of the proposed geometric decomposition, let us start by considering the flux of the average excess availability:
\begin{equation}\label{eq:change_ex}
    \D_t \langle \FF_{\ex} \rangle = \langle \vec{v}*\nabla\FF_{\ex}\rangle = \average{\vec{v}_{\ex}*\vec{v}_{\ex}} + \langle \vec{v}_{\hk}*\nabla\FF_{\ex}\rangle \;,
\end{equation}
where the first term on the r.h.s. can be defined as the excess dissipation:
\begin{equation}
    \label{eq:ex_dis}
    \dot{S}_{\ex} = \average{\vec{v}_{\ex}*\vec{v}_{\ex}}\geq 0 \;,
\end{equation}
while the last contribution is the flux of $\mathcal{F}_\ex$ along the direction of $\vec{v}_\hk$. From Eq.~\eqref{eq:change_ex} and the orthogonality condition in Eq.~\eqref{eq:gradient_perp}, at stationarity we have:
 \begin{eqnarray}\label{eq:flux_ex}
     \dot{S}^{\st}_{\ex} &=& -\average{ \vec{v}_{\hk}*\nabla\FF_{\ex}}^{\st} = - \left( \mathcal{F}_\ex \right)_{\flux,\hk}^\st \geq 0 \;.
 \end{eqnarray}
Similarly, we can define the housekeeping dissipation as:
\begin{equation}
    \label{eq:hk_dis}
   \dot{S}_{\hk} = \average{\vec{v}_{\hk}*\vec{v}_{\hk}}\geq 0 \;.
\end{equation}
By combining Eqs.~\eqref{eq:excess_time_dec}, \eqref{eq:hk_dis}, and \eqref{eq:ex_dis}, we can rewrite the total entropy production as:
\begin{equation}
\begin{split}
\dot{S}_{\tot}&=\dot{S}_{\hk}+\dot{S}_{\ex}+2\langle \vec{v}_{\hk}*\nabla \FF_{\ex}\rangle\\
    &=\dot{S}_{\hk}+d_t\langle \FF_{\ex}\rangle +\langle \vec{v}_{\hk}*\nabla \FF_{\ex}\rangle \;.
    \end{split}
\end{equation}
Hence, from Eq.~\eqref{eq:ex_dis}, we arrive to our first main result:
\begin{eqnarray}\label{eq:excess_house_keeping_bal}
     \dot{S}^{\st}_{\tot}=\dot{S}^{\st}_{\hk}-\dot{S}^{\st}_{\ex}
\end{eqnarray}
implying the following upper limits for the dissipation:
\begin{eqnarray}\label{eq:excess_house_keeping_limits}
    \dot{S}^{\st}_{\tot}\leq\dot{S}^{\st}_{\hk} \;, \qquad  \dot{S}^{\st}_{\ex}\leq\dot{S}^{\st}_{\hk} \;.
\end{eqnarray}
We are now in a position to propose a thermodynamic interpretation of this result. When driven away from equilibrium by external conditions, a system develops dissipative currents, while also potentially modifying its availability with respect to the equilibrium Boltzmann distribution. The housekeeping and excess part of the velocity quantify these two different contributions. 
The system is kept out of equilibrium, causing a non-vanishing $\vec{v}_{\rm hk}$. Indeed, if this contribution disappeared, the stationary velocity would have only gradient term and the system would relax to an effective equilibrium distribution. The presence of $\vec{v}_{\rm hk}$ results in an energy inflow from the environment which is balanced by the excess dissipation into the heat bath $\dot{S}_\ex^\st$, Eq.~\eqref{eq:flux_ex}. Hence, according to Eq.~\eqref{eq:excess_house_keeping_bal}, the net dissipation into the environment is given by the house-keeping part, due to non-equilibrium conditions, minus the excess dissipation which allows the system to have a wider distribution than at equilibrium, for instance by populating higher-energy states more than in the corresponding equilibrium scenario.

We remark that the proposed decomposition differs from those in \cite{dechant2022geometric,maes2014nonequilibrium}, whereas the one introduced in \cite{hatano2001steady} coincides with the adiabatic splitting considered in this work. The decompositions presented here are motivated by geometric and thermodynamic considerations, and the term ``excess'' has been adopted in analogy with previous studies on the Onsager excess dissipation functional, whose relevance will become clear later on.

\subsection{Non-equilibrium from non-conservative forces}\label{sec:non_cons_force}
A more transparent physical interpretation of the excess availability stems by the study of the simple case in which non-equilibrium conditions originate from the presence of non-conservative forces, $\vec{f}$ (e.g., external fields, shear flows). In this scenario:
\begin{eqnarray}
F^{i}=f^i-\partial^{i}U, \quad
 D^{ij}=\delta^{ij}T,\quad \nabla^{i}=T\partial^i
\end{eqnarray}
with $\partial^{i}=\delta^{ij}\partial_j$, $U$ is the potential acting on the system, and $T$ the temperature of the external bath. We also employ the Einstein relation to link diffusion and temperature to ensure that, when $\vec{f} = \vec{0}$, the system relaxes to the Boltzmann distribution. Here, $\beta = T^{-1}$ with the Boltzmann constant set to $1$ for simplicity.
From the force decomposition in Eq.~\eqref{eq:force_dec}, the adiabatic velocity can be expressed as follows:
\begin{eqnarray}
\vec{v}_{\rm a}&=&\vec{f}-\partial U+\nabla\mathcal{F} = \vec{f} + \nabla \left(\mathcal{F} - \beta U \right) \;,
\end{eqnarray}
where all gradient terms have been isolated and can be identified with $\nabla \mathcal{F}_{\ex,\A}$. The first contribution coincides with the non-conservative force - the sole source of non-equilibrium in this scenario - and corresponds to $\vec{v}_\hk$.  Thereby, in this simple scenario, we obtain a unique decomposition of the velocity:
\begin{eqnarray}
    \vec{v}_{\hk} = \vec{f} \;,\qquad \vec{v}_{\ex,\A} = \nabla \left(\mathcal{F} - \beta U \right) = \nabla\FF_{\ex,\A} \;,
\end{eqnarray}
with $\mathcal{F}_{\ex,\A} = \mathcal{F}_\ex$ at stationarity since the non-adiabatic (time-dependent) contribution vanishes. Out of equilibrium, the particle could visit excited states with availability larger then the equilibrium counterpart. Hence, we identify the deviation of $\FF$ from its equilibrium value at the stationary state with the excess availability.
The adiabatic entropy production is thus given by
\begin{eqnarray}
\dot{S}_{\A}=\langle \vec{v}_\A*\vec{v}_{\rm a}\rangle =\langle \vec{v}*\vec{f}\rangle+\langle \vec{v}*\nabla \mathcal{F}_{\ex,\A}\rangle,
\end{eqnarray}
where the first term corresponds to the heat dissipated by the non-conservative force, and the second one to the excess heat (and duly vanishes in the stationary state). In the steady state we have:
\begin{eqnarray}
\dot{S}^\st_{\rm a}&=&\langle \vec{v}_{\rm a} *\vec{f}\rangle_{\st}=\beta\Big(\underbrace{\langle \vec{f}^2\rangle_{\st}}_{\dot{Q}^{\rm out}} - \underbrace{\langle\nabla\mathcal{F}_{\ex}^2\rangle_{\st}}_{-\dot{Q}^{\rm in}}\Big)\geq 0 \;.
\end{eqnarray}
The first term multiplying the inverse temperature can be interpreted as the heat due to the non-conservative force released into the environment, while the second one as the heat absorbed from the non-equilibrium conditions that modifies the system availability.

As a consequence, it is immediate to notice that we can define an excess entropy as follows:
\begin{equation}\label{eq:ex_entropy}
    S_{\ex} = -\int \D x P(x)\log\frac{P(x)}{P_{\eq}(x)}=\average{\FF_{\ex}}.
\end{equation}
This is the relative entropy with respect to the equilibrium state. As such, it is coordinate and coarse-graining independent, resulting to be a well defined physical observable \cite{Polettini_2012, banavar2007maximumrelativeentropyprinciple, busiello2019entropy}. In this context, the equilibrium distribution acts as a \emph{prior} with which the NESS has to be compared. As we will see in the next section, different choices of the prior might lead to different identification of excess and housekeeping contributions, especially in complex cases in which more than one physical current is present. This is somehow similar to what happens in electromagnetism, where a gauge-potential must be fixed (we will comment more on this point in the conclusions).
Finally, by computing the temporal derivative of the excess entropy, in analogy to what is done for $\dot{S}_{\rm sys}$ in Eq.~\eqref{eq:entropy_flux}, we connected this change to the excess dissipation as follows:
\begin{eqnarray}
    d_t   S_{\ex}=\dot{S}_{\ex}+\average{\vec{v}_{\rm hk}*\nabla \FF_{\rm ex}} \;.
\end{eqnarray}

\subsection{Non-equilibrium from multiple baths}
\label{sec:baths}

Another paradigmatic yet more intricate scenario to induce non-equilibrium conditions involves systems coupled to multiple baths, such as the Brownian gyrator and a Brownian particle in contact with two thermal reservoirs. In these cases  
the identification of $\mathcal{F}_\ex$ becomes more subtle since there are multiple possible equilibrium distributions, i.e., one for each bath, $P_{\eq}^{T_i} \propto \exp(-U/T_i)$, where $U$ is the potential. However, interaction terms in $U$ between different components might result into an interaction between different baths. Indeed, in general $U = \sum_{i}U_i+U_{\mathrm{int}}$, where $U_i$ depends only on the $i$-th component.
For simplicity, we assume that each degree of freedom $i$ is coupled to a different bath at temperature $T_i$, i.e., $D^{ij}=\delta^{ij} \sqrt{T_{i}T_{j}}$.
The stationary velocity for each component can be readily obtained as before:
\begin{equation}
  v^i_{\st} = - \partial^i U + \nabla^i \mathcal{F} = - \partial^i U_{\rm int} + \nabla^i \left( \FF - \frac{U_i}{T_i} \right) \;.
\end{equation}
The second term is a gradient and, as such, can be identified with the contribution from the excess availability. Indeed, it accounts for the difference between $\FF$ and the equilibrium case, where no interaction exists between components and each one relaxes to the Boltzmann distribution associated with its respective bath. As a consequence, the first term coincides with the house-keeping velocity and, once again, only contains external contributions triggering non-equilibrium conditions. Thus:
\begin{eqnarray}\label{eq:dec_multiple_bath}
\vec{v}_{\hk} = - \partial U_{\mathrm{int}} \;, \qquad
\vec{v}_{\ex,\A} = \nabla \Bigg( \underbrace{\FF - \sum_i \frac{U_i}{T_i}}_{\FF_{\ex,\A}} \Bigg) \;.
\end{eqnarray}
In this case, it is manifest how the choice of the reference equilibrium distribution changes the identification of the housekeeping and excess contributions. Another possibility would have been to compare the NESS with the equilibrium distribution where $T_i = T$ for all baths, in the presence of an interaction potential. In such a case, the component of the velocity encoding the external non-equilibrium conditions would have encoded the fact that baths have different temperatures.
To conclude this case, we note that the metric is flat but anisotropic, with $D^{-1} = \operatorname{diag}(1/T_1, 1/T_2, \dots, 1/T_d)$. It can be mapped to the standard Euclidean metric by rescaling each degree of freedom by $1/\sqrt{T_i}$. Thermodynamically, this implies that in this rescaled space, each direction is effectively ``measured'' by a different temperature, i.e., the length-scale along direction $i$ is inversely proportional to its temperature $T_i$. We also generalize our framework to include multiple thermal baths acting simultaneously on the same degrees of freedom and generating different currents. In this case, since there is not only one physical current at play, Eq.~\eqref{eq:db1} does not coincide with the equilibrium condition, hence the reference equilibrium distribution does not naturally appear in the stationary velocity and has to be set a priori. As a concrete example, and for simplicity of notations, we consider a one-dimensional Brownian particle attached to two baths at different temperatures and subject to two corresponding potentials \cite{parrondo_ratchet,van2010three}:
\begin{eqnarray}
    \dot{x} = - \partial_x U_1 + \sqrt{2 T_1} \xi_1(t) - \partial_x U_2 + \sqrt{2 T_2} \xi_2(t) \;.
\end{eqnarray}
The system reaches a generalized Boltzmann distribution with an effective temperature:
\begin{eqnarray}
    P_{\st} \propto e^{-\FF}, \qquad \FF=\frac{U_1+U_2}{T_1+T_2} \;.
\end{eqnarray}
In this particular condition, the total current is zero, as the contributions from each bath balance each other. Nevertheless, as we shall see shortly, there is an ongoing entropy production. Moreover, the absence of interactions (or any other external non-equilibrium source) implies that $v_{\rm a}$ stems purely from excess terms from a geometrical perspective:
\begin{equation}
\begin{split}
    v_{\rm a} &= v_{1,\ex} + v_{2,\ex} = 0 \;, \\
    v_{\alpha,\ex} &= - \partial_x U_\alpha + T_\alpha \partial_x \FF \qquad \alpha = 1,2 \;. 
    \end{split}
\end{equation}
However, when we set a reference equilibrium distribution, say $P_{1}\sim \exp(-U_1/T_1)$, the presence of the second bath can be identified as the non-equilibrium driving of the system and, as such, it has to reflect into the housekeeping part of velocity:
\begin{eqnarray}
    v_{\ex,\A}=v_{1,\ex},\quad v_{\rm hk}=v_{2,\ex}.
\end{eqnarray}
Hence, the total dissipation is given by the excess contribution associated with thermal conduction:
\begin{eqnarray}
    \dot{S}_{\rm a}=\dot{S}_{\rm ex} = \sum^{2}_{i=1}\average{\frac{v_{i,\ex}^2}{T_i}} = \average{\frac{v_{\ex}^2}{T_1}}+\average{\frac{v_{\hk}^2}{T_2}}\;.
\end{eqnarray}
We emphasize here that a purely geometrical interpretation of $v$ would give only excess contributions. However, a physically informed interpretation of our results in the presence of multiple currents, based on Eq.~\eqref{eq:ex_entropy}, leads to one of the possible correct housekeeping/excess decompositions (depending on the equilibrium prior).

\section{Towards a NESS: dissipation and symmetry breaking}\label{sec:dyn}

\subsection{Revisiting  Prigogine and Onsager principles}\label{sec:onsager_prigogine1}
We have presented how the geometric properties of the stationary velocity reflect non-trivial thermodynamic properties of the NESS. Now, we highlight how such a non-equilibrium state is reached. To answer to this question, we consider the Kullback-Leibler divergence between $P(\vec{x},t)$ and the stationary distribution $P_{\rm st}(\vec{x})$ that is known to be a Lyapunov function of the dynamics \cite{schnakenberg1976network}. It is defined as: 
\begin{eqnarray}
    D_{\KL}(t) &=& \int d\vec{x} ~P(\vec{x},t) \log \left( \frac{P(\vec{x},t)}{P_{\rm st}(\vec{x})} \right) \nonumber \\
    &=& \average{\FF_{\ex,\A}} - S_{\ex}= - \langle \FF_{\ex,\rm na}\rangle \;.
\end{eqnarray}
where  equalities follow from Eq.~\eqref{eq:ex_entropy}. Its time-derivative has to be non-positive, vanishes at stationarity, and is equal to the non-adiabatic entropy production: 
\begin{eqnarray}
    -\D_t D_{\KL}(t) = \dot{S}_{\na} \geq 0 \;.
\end{eqnarray}
By employing the adiabatic and the housekeeping decomposition in Eqs.~\eqref{eq:entropy_adiabatic} and \eqref{eq:dec_hk_ex_a}:
\begin{eqnarray}\label{eq:Gfunc}
     -\D_t D_{\KL}(t) &=& \dot{S}_{\na}=\dot{S}_{\tot}-\dot{S}_{\A} \\
     &=&\underbrace{\dot{S}_{\tot}-\langle \vec{v}*\vec{v}_{\hk}\rangle-\langle \vec{v}*\nabla\FF_{\ex}\rangle}_{\mathcal{G}_\na} \geq 0 \nonumber \;,
\end{eqnarray}
where $\mathcal{G}_\na$ is a functional that is minimized when $P = P_\st$ and decreases during the system evolution towards stationarity. Its expression reveals how the geometrical properties of $\vec{v}_{\rm a}$ enter the game. To have a clear physical understanding of the result, let us first note that in an equilibrium condition the criterion trivially corresponds to the minimum entropy production, or entropy maximization. Away from equilibrium, Eq.~\eqref{eq:Gfunc} states that, while reaching a NESS, the system has a non-negative non-adiabatic entropy production, equivalent to the Glansdorff-Prigogine criterion. 
Given that in non-equilibrium conditions both total and adiabatic contributions are positive, Eq.~\eqref{eq:Gfunc} can be interpreted as the tendency to minimize the entropy production compatibly with maximizing the adiabatic dissipation, composed by the sum of the house-keeping and excess fluxes, i.e., aligning $\vec{v}$ with $\vec{v}_{\A}$.
Once a NESS is reached, we would like to understand its associated geometric properties of the adiabatic velocity as resulting from a general variational principle. To do so, we construct a functional which is reminiscent of the Onsager's excess dissipation functional, recently generalized to non-equilibrium conditions in the context of MFT \cite{Onsager1, MFT, Bertini2004MinimumDP}. From Eq.(\ref{eq:Gfunc}), it is clear that the only part of the velocity contributing to the (unavoidable) stationary dissipation is $\vec{v}_{\hk}$. Hence, we consider that the velocity would try to minimize the excess dissipation, i.e., those not stemming from $\vec{v}_{\hk}$, with the constraint to satisfy the orthogonality condition, Eq.~\eqref{eq:gradient_perp}. To implement this principle, we construct a variational functional $\mathcal{G}_{\ex}$ for a generic velocity field $\vec{v}(x)$ with a given $P(\vec{x})$ implementing the constraint via a Lagrange multiplier:
\begin{equation}
\begin{split}
    \mathcal{G}_{\ex}&=\frac{1}{2}\average{\left(\vec{v}-\vec{v}_{\hk}\right)*\left(\vec{v}-\vec{v}_{\hk}\right)}-\lambda\average{ \vec{v}*\nabla\mathcal{F}_{\ex,\rm a}} \\
    &=\frac{1}{2}\dot{S}_{\ex}-\lambda d_t \average{\FF_{\ex, \rm a}} \;.
   \label{eq:ex_onsager}
    \end{split}
\end{equation}
where $\lambda$ is the multiplier to be determined, 
Considering $P(\vec{x})$ as fixed, by taking the variation with respect to $\vec{v}$ to zero   yields the optimal value $\bar{v}$:
\begin{eqnarray}\label{eq:ex_min}
    \delta_{\vec{v}} \mathcal{G}_{\ex}&=&0 \; \Leftrightarrow \; \vec{\bar{v}}=\vec{v}_{\hk}+\lambda\nabla\FF_{\ex,\rm a}.
\end{eqnarray}
The derivative with respect to $\lambda$ gives the constraint:
\begin{eqnarray}
       \partial_{\lambda} \mathcal{G}_{\ex}=\average{\vec{v}*\nabla\FF_{\ex,\A}}=0
\end{eqnarray}
that when evaluated at the extremal values $\bar{\vec{v}}$ and $\bar{\lambda}$ gives
\begin{eqnarray*}
   \partial_{\lambda}\mathcal{G}_{\ex}({\vec{\bar{v}}})= \average{\vec{\bar{v}}*\nabla\FF_{\ex,\A}}=0 \;\Leftrightarrow \; \bar{\lambda}=-\frac{ \average{\vec{v}_{\hk}*\nabla\FF_{\ex,\A}}}{\dot{S}_{\ex,\A}} \;.
\end{eqnarray*}
with the adiabatic excess entropy defined as
\begin{equation}\label{eq:ex_a_dis}
    \dot{S}_{\rm ex,a}=\average{\nabla\FF_{\rm ex,a}*\nabla\FF_{\rm ex,a}}\geq 0
\end{equation}
in analogy with  Eq.~\eqref{eq:ex_dis}. By evaluating the functional in the minimum  $(\bar{\vec{v}},\bar{\lambda})$, we obtain $\mathcal{G}_{\ex}$ as
\begin{eqnarray}
    \mathcal{G}_{\ex}(\bar{\vec{v}},\bar{\lambda})=\frac{1}{2}\frac{\average{\vec{v}_{\hk}*\nabla\FF_{\ex,\A}}^2}{\dot{S}_{\ex}}\geq 0 \;,
\end{eqnarray}
which, when evaluated at the correct stationary state $-\log P=\FF$ gives:
\begin{eqnarray}
\bar{\lambda}=1, \quad\bar{\vec{v}}&=&\vec{v}_{\A}=\vec{v}_{\hk}+\nabla\FF_{\ex,\A},\quad  \;,
\end{eqnarray}
is equal to (half) the adiabatic excess entropy production, i.e., one has
\begin{eqnarray}
\mathcal{G}_{\ex}^{\st}(\bar{\vec{v}},\bar{\lambda})=\frac{1}{2}\dot{S}^{\st}_{\ex,\A} \;.
\end{eqnarray}
Hence, the correct stationary velocity minimizes $\mathcal{G}_\ex$ while satisfying the  orthogonality condition.

Furthermore, from the results of Sec.~\ref{sec_:ex_hk}, $\bar{\lambda}$ can be interpreted as the ratio between the input and output excess heat. These contributions must balance each other at stationarity, so that $\bar{\lambda} = 1$. Notice also that, while Eq.~\eqref{eq:ex_onsager} is inspired by the Onsager  quadratic excess dissipation functional, at variance with it, it determines just the stationary velocity and not the full dynamics. Indeed, Eq.~\eqref{eq:ex_onsager} is not an explicit time derivative and is not bound to be a monotonic function of time. This is true just when detailed balance is satisfied, where 
\begin{eqnarray}
   -\D_t D_{\rm KL}(t) = \mathcal{G}_{\na} = 2~\mathcal{G}_{\ex} = \dot{S}_{\tot} \;,
\end{eqnarray}
and the dynamics is equivalent to a variational minimization of the dissipation. Finally, Eq.~\eqref{eq:ex_onsager} can be rewritten in a way that resembles Eq.~\eqref{eq:Gfunc}:
\begin{equation}\label{eq:G_ex_align}
\begin{split}
     \mathcal{G}_{\ex}&=\frac{1}{2}\average{\vec{v}*\vec{v}}+\frac{1}{2}\average{\vec{v}_{\hk}*\vec{v}_{\hk}}\\
     &\qquad\qquad {}-\average{\vec{v}*(\vec{v}_{hk}+\lambda\nabla\FF_{\ex,\A})} \;.
     \end{split}
\end{equation}
This expression clarifies that minimizing $\mathcal{G}_{\ex}$ with respect to $\vec{v}$ amounts to minimizing the velocity modulus, i.e., the total entropy production, while maximizing the alignment between $\vec{v}$ and $\vec{v}_{\A}$, coherently with the intuition stemming from Eq.~\eqref{eq:Gfunc}. The component $\vec{v}_{\A}$ seems to act as an ``external field'' on the system, imposing a symmetry breaking along its geometry. While $\vec{v}_{\hk}$ can be thought as real external field stemming from the external conditions, $\vec{v}_{\ex}$ emerges from the system itself. Nevertheless, at the end of this section, we show that the excess availability can be directly derived by averaging over stochastic trajectories.

\subsection{Symmetry breaking in the trajectory space}

In our framework, we can improve the physical interpretation of the principles in Eqs.~\eqref{eq:ex_onsager} and \eqref{eq:Gfunc} to reveal their relationship with symmetry breaking phenomena in the trajectory space.  
The first term in Eq.~\eqref{eq:Gfunc} only dictates that the system tends to minimize its total dissipation, as for equilibrium relaxation phenomena. The second contribution emerges only out of equilibrium, hence encoding an additional dissipation stemming from the properties of the adiabatic velocity, and coincides with the entropy fluxes associated with the housekeeping and excess part of the force. 
Here, we show that it accounts for a tendency to circulate, breaking a parity symmetry holding at equilibrium and leading to a stationary dissipation into the environment. Considering a system governed by Eq.~\eqref{eq:Langevin}, the probability of observing a trajectory $\Gamma$ -- starting in $\vec{x}_0$ at $t=0$ and arriving in $\vec{x}_{\rm f}$ - is given by
\begin{equation}
    P(\Gamma) = \underbrace{e^{-\mathcal{A}(\Gamma)/2}}_{P(\Gamma|x_0)} P(x_0)
\end{equation}
where $\mathcal{A}(\Gamma)$ is the Onsager-Machlup action \cite{cugliandolo2017rules} which, using the Stratonovich prescription, reads
\begin{equation}\label{eq:onsager_action}
    \mathcal{A}(\Gamma) = \int_0^t d\tau \left[ \frac{1}{2} \left( \dot{\vec{x}} - \vec{F} \right)* \left(\dot{\vec{x}} - \vec{F}\right) + \nabla* \vec{F} \right] \;,
\end{equation}
in which all quantities have to be evaluated along $\Gamma$. 
Eq.~\eqref{eq:onsager_action} can be rewritten as the time integral of a lagrangian, $A$, and it can be divided into time-symmetric $A_{\rm s}$, stemming from an ``energy-stress tensor'',  and time-antisymmetric parts,  $A_{\rm a}$ :
\begin{eqnarray}
   \mathcal{A}(\Gamma) &=& \int_0^t d\tau A,\quad 
   A=D_{ij}A_{\rm s}^{ij}+D_{ij}A_{\rm a}^{ij}\\
  A_{\rm s}^{ij} &=&\frac{1}{2}\dot{x}^i\dot{x}^j+\frac{1}{2}F^iF^j+\partial_k D^{ik}F^j\\
  A_{\rm a}^{ij} &=&-\dot{x}^iF^{j}.
\end{eqnarray}
The detailed balance condition in Eq.~\eqref{eq:db1} can thus be rewritten as a condition for the stress-energy tensor to be symmetric with respect to indices:
\begin{eqnarray}\label{eq:asym_tensor}
   w_{ij} &=& (A_{\rm s})_{ij}-(A_{\rm s})_{ji} = \partial_{i} F_{j}-\partial_{j} F_{i} \nonumber \\
   &=&\partial_{i} (v_{\hk})_j - \partial_{j} (v_{\hk})_i \;,
\end{eqnarray}
where $w_{ij}$ is the ``torque'' tensor, i.e., the antisymmetric part of the stress-energy tensor. This field encodes the non-conservation of the angular momentum, hence parity symmetry breaking, and coincide with the vorticity field $\vec{w}$ in $2$ and $3$ dimensions. The dissipation along a trajectory $\Gamma$, $s(\Gamma)$, customarily quantified by the ratio between the probability of $\Gamma$ and its time reversal $\tilde{\Gamma}$ \cite{seifert2012stochastic}, is associated with the antisymmetric part of the action. To deepen our understanding, consider that in a closed trajectory $\Gamma_\circ$, i.e., $x_{0}=x_{\rm f}$, we have:
\begin{equation}
    s(\Gamma_\circ) = \log\frac{P(\Gamma_\circ)}{\tilde{P}(\tilde{\Gamma}_\circ)} = \int_{\Gamma_\circ} d\vec{x} * \vec{v}_{\hk} \equiv s_{\hk}(\Gamma_\circ) \;.
\end{equation}
In $2$ and $3$ dimensions, we can further apply Stokes' theorem and thus rewrite the dissipation as the flux of the vorticity field $\vec{w}$ across the surface $\vec{\Sigma}(\Gamma_\circ)$ defined by the closed trajectory:
\begin{eqnarray}
    s_{\hk}(\Gamma_\circ) = \int_{\vec{\Sigma}(\Gamma_\circ)} d\vec{\Sigma} ~\vec{w} \;.
\end{eqnarray}
These results establish a clear connection between the propensity of performing closed trajectories in a preferential direction, i.e., the breaking of the equilibrium parity symmetry, and the house-keeping velocity that breaks the space symmetries in the stress-energy tensor. Furthermore, these observations reinforce the validity of our identification of housekeeping contribution as stemming from the non-equilibrium conditions, and the excess availability $\mathcal{F}_\ex$ as capturing pseudo-equilibrium modifications. Indeed, we notice that $\mathcal{F}_\ex$ does not contribute to the dissipation along any closed trajectory $\Gamma_\circ$.

Furthermore, if we look at the dissipation along an open trajectory, we note that the gradient part of $\vec{v}_\A$ can give an additional contribution proportional to the initial and final state, indicating an additional directional preference:
\begin{eqnarray}
\label{eq:eq_start}
\log\frac{P_{\eq}(\vec{x}_0)P(\Gamma|x_0)}{P_{\eq}(\vec{x}_f)\tilde{P}(\tilde{\Gamma}|x_f)} &=& 
s_{\hk}(\Gamma) \\
\label{eq:open_traj}
\log\frac{P_{\st}(\vec{x}_0)P(\Gamma|x_0)}{P_{\st}(\vec{x}_f)\tilde{P}(\tilde{\Gamma}|x_f)} &=& 
s_{\hk}(\Gamma)+\Delta \mathcal{F}_\ex
\end{eqnarray}
where $\Delta \FF_{\ex} = \FF_\ex(\vec{x}_f) - \FF_\ex(\vec{x}_0)$ and $\mathcal{F}_\ex = \mathcal{F} - \beta U$. Eq.~\eqref{eq:open_traj} explains the role on single trajectories of the housekeeping and excess velocities: The first one selects the probability of producing a fluctuation  from the equilibrium state over its time reversal, while for the equivalent quantity in a NESS also the excess is necessary.
Indeed, it tends to select a trajectory over its time-reversed if it leads to a increase in excess availability 
When averaged over trajectories, initial, and final states \cite{seifert_entropy,busiello2020entropy,seifert2012stochastic}, the two contributions on the rhs of Eq.~\eqref{eq:open_traj} lead exactly the last two terms of Eq.~\eqref{eq:Gfunc}. Hence, when approaching the NESS, the system minimizes the total dissipation compatibly with maximizing the breaking of two symmetries in the trajectory space: one related to the direction of rotation and associated with the house-keeping velocity; the other determining the phase-space occupation imposed by non-equilibrium conditions and connected to the excess availability flux. These findings reveal the profound physical meaning of the main results of this work, also providing a broader context for recent results on a topological fluctuation theorem \cite{mahault2022topological} and gauge symmetries in thermodynamics \cite{Polettini_2012,polettini_dice,wang_gauge}.

\subsection{A variational principle for NESS and trajectory-dependent estimation of $\FF_\ex$}
Following our findings, we can build a variational principle to derive the NESS distribution that only requires some information on single-trajectory thermodynamics. This result resembles a non-equilibrium version of the maximum entropy approach for equilibrium systems \cite{jaynes}.
We propose the Jarzynski statistics of the house-keeping entropy production over trajectories:
 \begin{eqnarray}
     \theta_{\hk}(x)=
     \log\langle e^{-s_{\hk}(\Gamma)}\rangle^{\eq}_{\Gamma}(x),
 \end{eqnarray}
that is exponential of minus the housekeeping entropy production averaged over all trajectories starting at $x$ and lasting till $t$, weighted by the equilibrium distribution $\langle \cdot \rangle^{\eq}_{\Gamma}(x)=\int \D\Gamma \cdot P(\Gamma) P_{\eq}(x)$. The excess entropy functional subjected to a constrain that imposes a fixed value $\Theta$ for the average of $\theta_{\hk}$ is
\begin{eqnarray}\label{eq:max_ent}
    \mathcal{L}_{\ex}=S_{\ex}-\lambda\left(\langle \theta_{\hk}\rangle-\Theta\right),
\end{eqnarray}
where $\lambda$ is the Lagrange multiplier. The maximization of Eq.~\eqref{eq:max_ent} gives
 \begin{equation}
     P(x) = P_{\eq}(x)e^{-\theta_{hk}(x)} \;, \qquad
     \FF_{\ex}(x) = \theta_{hk}(x) \;,
     \label{eq:Fex_traj}
\end{equation}
consistently with past and recent results in non-equilibrium physics \cite{kawasaki,  mclennan1989introduction, komatstu_ness, Colangeli_2011}. Equivalently to Eq.~\eqref{eq:Gfunc}---whose writing requires prior knowledge of the steady-state distribution---this variational principle tells that the stationary state is the one of maximum relative entropy compatible with a given adiabatic excess availability. Crucially, Eq.~\eqref{eq:Fex_traj} provides an operative way to quantify the excess availability from an average of the housekeeping dissipation over trajectories.

\section{Geometric characterization of non-equilibrium steady states} \label{sec:NESS}
\begin{table*}[t]
\renewcommand{\arraystretch}{1.6}
\begin{tabular*}{\textwidth}{@{\extracolsep{\fill}} lcccc}
\toprule
\; & \textbf{Open} & \textbf{Locally closed} & \textbf{Locally balanced} & \textbf{Local equilibrium} \\
\midrule
Availability & $\nabla * \vec{v}_{\A} = \vec{v}_{\A} * \nabla \FF$ & $\vec{v}_{\A} * \nabla \FF = 0$ & $\nabla^2 \FF_{\ex,\A} = 0$ & Same as locally balanced \\
Velocity & $\vec{v}_\A = \vec{v}_\hk + \vec{v}_{\ex, \A}$ & $\nabla * \vec{v}_\hk = - \nabla * \vec{v}_{\ex,\A}$ & $\nabla * \vec{v}_\hk = 0$ & Same as locally balanced \\
Extreme points & $\nabla \FF^* = 0$ & $\vec{v}_{\hk}^* = -\vec{v}_{\ex,\mathrm{a}}^* = \beta \nabla U^*$ & Same as locally closed & $\nabla \FF^* = \beta \nabla U^* = 0$ \\
\hline
Noise expansion & $P_{\st} \sim e^{-\FF^{(0)}/\sigma - \FF^{(1)}}$ & $P_{\st} \asymp e^{-\FF^{(0)}/\sigma}$ & Same as locally closed & $P_{\st} = Z^{-1} e^{-\FF^{(0)}/\sigma}$ \\
\bottomrule
\end{tabular*}
\caption{\textbf{NESS characterization}. Classes toward the right require increasingly stringent conditions. The first three rows determine the geometric characterization discussed in Sec.~\ref{sec:NESS}, while the last row indicate noise expansion properties (Sec.~\ref{sec:noise_exp}).}
\label{ness_ch}
\end{table*}

We propose here a NESS characterization scheme which is based on the excess-housekeeping decomposition of the velocity and its geometric properties. We introduce and discuss the concepts of \emph{locally} closed, locally balanced, and local equilibrium systems. In Table \ref{ness_ch}, we report a brief summary of this classification.

Let us start from Eq.~\eqref{eq:stationary} that dictates the relationship between $\vec{v}_\A$ and $\mathcal{F}$ in a generic NESS. A relevant class of non-equilibrium systems, however, satisfies a stringent condition on the velocity: 
\begin{eqnarray}
\label{eq:closness}
\nabla*\vec{v}_{\A}=0=\vec{v}_{\A}*\nabla\FF.
\end{eqnarray}
where the last equality comes from Eq.~\eqref{eq:stationary}. Therefore, the adiabatic velocity must be tangent to the contour lines of the availability, so that the NESS is characterized by closed fluxes in the configuration space. We name these systems locally closed, because the stationary probability does not undergo local expansion or contraction along the velocity field, in direct analogy with incompressible flow in fluid dynamics (see Eq.~\eqref{eq:fp_div}). Eq.~\eqref{eq:closness} further implies that:
\begin{equation}
    \label{eq:closeness}
    \nabla*\vec{v}_{\hk} = - \nabla*\vec{v}_{\ex, \rm a} \;.
\end{equation}
Moreover, again from Eq.~\eqref{eq:closness}, we have that the adiabatic velocity must be zero at the extreme points of the availability, i.e., where $P_\st$ is maximized:
\begin{eqnarray}
    \nabla\FF(\vec{x}^*) \equiv \nabla\FF^* = 0 \rightarrow \vec{v}_{\A}(\vec{x}^*) \equiv \vec{v}^*_\A = 0 \;,
\end{eqnarray}
with the short-hand notation $X^* = X(\vec{x}^*)$. Hence, for locally closed systems, both the house-keeping and excess velocities and their fluxes need to balance each other at $\vec{x}^*$, implying that the house-keeping contribution is also equal to the local change in energy:
\begin{eqnarray}\label{eq:closed_condition_ex}
    -\vec{v}^*_{\hk} &=& \vec{v}^*_{\ex, \rm a} = \nabla\FF_{\ex, \A}^*=-\beta \nabla U^* \;.
\end{eqnarray}
The conditions in Eqs.~\eqref{eq:closed_condition_ex} and \eqref{eq:closeness} are very general and state that any source for the housekeeping flow must be a sink for the excess one and vice-versa, together with the fact that they must have common center/saddles. Furthermore, let us note that, by using Eq.~\eqref{eq:dec_hk_ex_a}, Eq.~\eqref{eq:closness} is equivalent to a Poisson equation for the excess availability where the divergence of the housekeeping velocity acts as the source term \cite{maes_poisson}:
\begin{eqnarray}\label{eq:possion-non-eq}
    \nabla^2\FF_{\ex,\rm a}=-\nabla*\vec{v}_{\hk} 
\end{eqnarray}
To interpret Eq.~\eqref{eq:possion-non-eq}, we recall that, intuitively, the Laplacian measures the local concavity of field, or, more rigorously, the difference between the field at that point and its average in a local neighborhood \cite{laplacian}. Hence, how much the availability concavity deviates from equilibrium is directly influenced by the housekeeping velocity that reflects the external non-equilibrium constrains.

Nevertheless, the condition defining a locally closed system contains the great majority of thermodynamic models. To narrow down the space of possible geometric properties, a natural simplification from Eq.~\eqref{eq:possion-non-eq} is to consider the additional constraint:
\begin{eqnarray}\label{eq:locally_balance}
    \nabla^2\FF_{\ex, \A}=0=\nabla*v_{\hk}.
\end{eqnarray}
This condition amounts to assuming that the excess availability is harmonic in the entire phase-space. As a consequence, the equilibrium and non-equilibrium availability have the same Laplacian:
\begin{eqnarray}\label{eq:harmonic_ex}
    \nabla^2\FF_{\ex,\A}=0\rightarrow \nabla^2\FF=\beta \nabla^2U.
\end{eqnarray}
We name this class of systems \emph{locally balanced}.
Finally, as a minimal way of constructing a NESS, we consider an even more stringent requirement that we name local equilibrium. In addition to local balance, we impose that the equilibrium and non-equilibrium availabilities have the same extreme point, $\vec{x}^*$ 
\cite{Qian_2015,mendler2020predicting}:
\begin{eqnarray}\label{eq:local_eq1}
\nabla\FF^*=\beta \nabla U^*=0 \;.
\end{eqnarray}
It is trivial to show that these conditions are valid for any linear system, i.e., with $\vec{F}=A\vec{x}$. From Eq.~\eqref{eq:local_eq1}: 
\begin{eqnarray}\label{eq:vel_local_eq}
    \vec{v}^*_{\ex,\A} = \vec{v}^*_{\hk} = 0,
\end{eqnarray}
The combination of Eqs.~\eqref{eq:locally_balance} and \eqref{eq:local_eq1} implies that the mean value of the excess availability in any configuration space region containing an extreme point must be zero. 
In synthesis, in local equilibrium systems, the geometry of the velocities is fully captured by a solenoidal house-keeping and an harmonic excess contribution:
\begin{eqnarray}\label{eq:dec_quasi_eq}
    (v_{\hk})^i&=&\partial_{j}B^{ij} \;, \quad B^{ij}=-B^{ji},\\
    (v_{\ex, \A})^i&=&D^{ij}\partial_j\FF_{\ex, \A} \;,\quad \partial_{i}D^{ij}\partial_i\FF_{\ex, \A}=0 \;.
\end{eqnarray}
Eqs.~\eqref{eq:dec_quasi_eq} and \eqref{eq:local_eq1} together signal that local equilibrium conditions support the existence of a current, induced by external conditions, that breaks rotational symmetry, while not allowing the system to stably occupy excited states at stationarity. Indeed, as we will show in the next section through two examples, local equilibrium systems can either occupy the same state space as in equilibrium, i.e., $S_{\ex, \A}=0$, that we hence define as \emph{pseudo-equilibrium}, or explore excited states via non-equilibrium fluctuations while having the same minimum as in equilibrium, i.e., $S_{\ex, \A}>0$.

Here, we present two pedagogical examples where the parity symmetry breaking stemming from the housekeeping dissipation is easy to visualize, and the possibility to explore excited states can be appreciated through the exact computation of the excess availability.

\subsection{Examples}
\begin{figure}[t]
    \centering
    \includegraphics[width=\columnwidth]{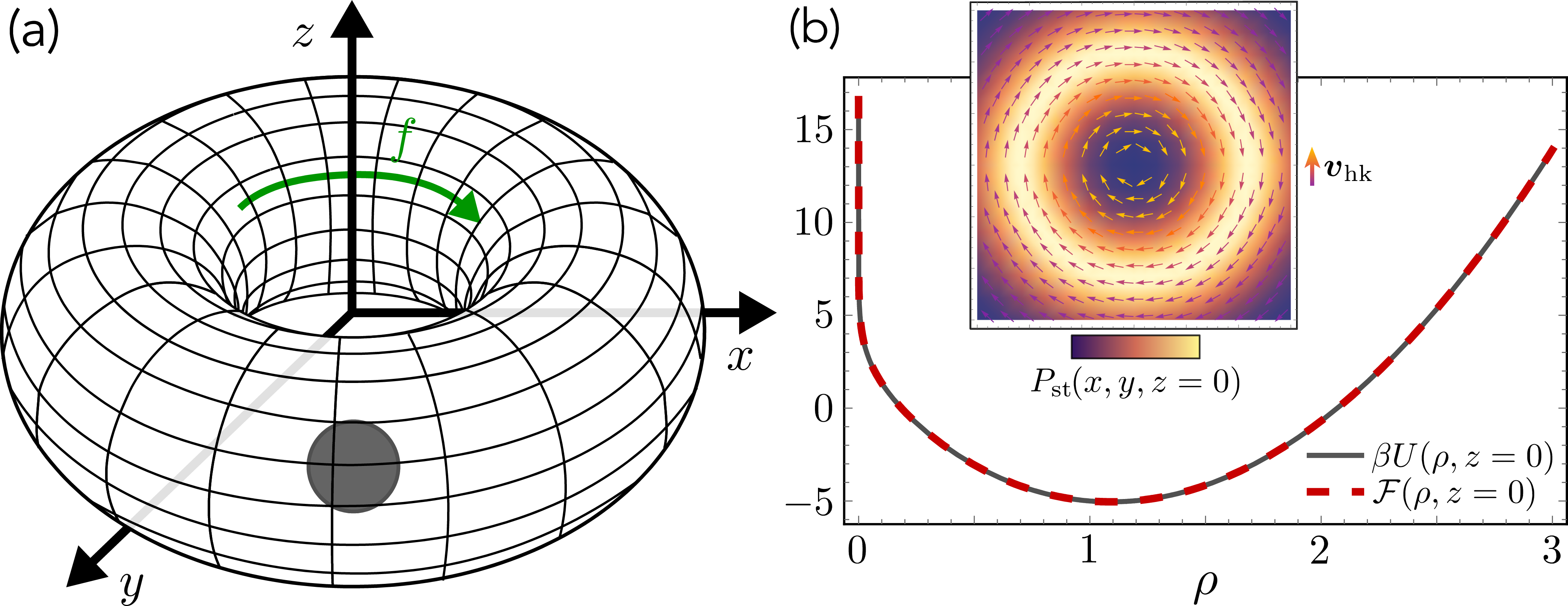}
    \caption{\textbf{Parity symmetry breaking for a driven particle in a circular potential.} (a) A sketch of the model. (b) In the NESS, the stationary availability (red dashed curve) coincides with the equilibrium potential (gray curve). Hence, the total stationary entropy production coincides with the housekeeping one, $\dot{S}^{\st}_{\rm tot} = \dot{S}^{\st}_{\rm hk}$. The inset show the pdf projected in the $(x,y)$ plane (in color-scale) with the velocity vector field on top (a bigger arrow corresponds to a stronger field). It clearly shows the tendency to rotate in a preferential direction, therefore breaking parity symmetry.}
    \label{fig:torus}
\end{figure}

\begin{figure*}[t]
    \centering
    \includegraphics[width=2\columnwidth]{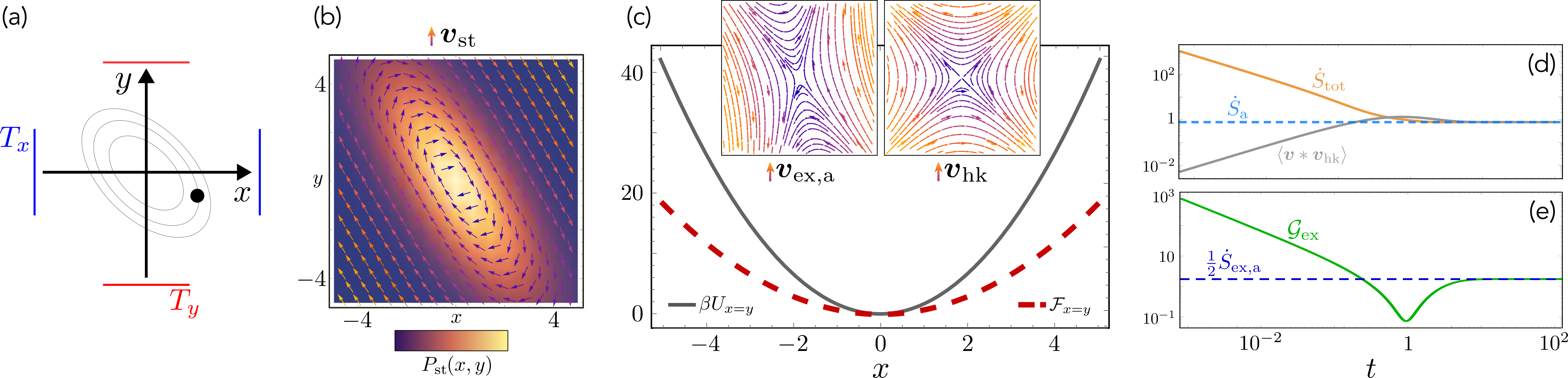}
    \caption{\textbf{Parity symmetry breaking in a Brownian gyrator.} 
    (a) A sketch of the model. (b) Steady-state distribution (in color-code) with the adiabatic velocity field on top (larger brighter arrows indicate a stronger field). (c) Availability (red dashed curve) and potential (gray curve) along the line $x=y$ show that excited states can be explored solely due to non-equilibrium fluctuations. In the inset, excess and house-keeping velocity at steady-state are presented in color-code and as a stream plot. (d) As a function of time, we show that the system tends to minimize $\dot{S}_{\rm tot}$ until reaching a non-zero value $\dot{S}_\A$ signaling a parity symmetry breaking. At the same time, the housekeeping flux $\langle \vec{v} * \vec{v}_\hk\rangle$ is the only surviving term at stationarity, with the excess availability flux going to zero (not shown). (e) The excess dissipation functional $\mathcal{G}_\ex$ is plotted, showing that it converges to $(1/2)\dot{S}_{\ex}$ at the steady state without being necessarily minimized along the dynamics.}
    \label{fig:gyrator}
\end{figure*}

\subsubsection{Driven particle in a circular potential}
Consider a $3D$ Brownian particle confined on a circle of radius $\rho^*$ by a (quadratic) potential, and driven along the circle itself by a constant non-conservative force $f$ (see Fig.~\ref{fig:torus}a).
It is instructive to write the Langevin equations describing this system in cylindrical coordinates $(\rho,\phi,z)$:
\begin{eqnarray}
\dot{\rho} &=& - \gamma (\rho - \rho^*) + \frac{D}{\rho} + \sqrt{2 D} ~\xi_\rho (t) \nonumber \\
\dot{\phi} &=& \frac{f}{\rho} + \frac{\sqrt{2 D}}{\rho} \xi_\phi (t) \\
\dot{z} &=& - \gamma z + \sqrt{2 D} ~\xi_z (t) \nonumber
\end{eqnarray}
with the equilibrium potential $U = (\gamma/2) (\rho - \rho^*)^2 + (\gamma/2) z^2$, the non-conservative force $f$ only acting in the direction of increasing the angle $\phi$, and $D/\rho$ appearing due to the Ito's formula for the change of variables. The stationary solution is found to support a probability flux only along the angle variable, $\phi$ but,
surprisingly, the steady-state availability coincides with the equilibrium potential (see Fig.~\ref{fig:torus}b). Hence, the excess availability is zero and the stationary velocity $\vec{v}_{\rm a} = \vec{v}_{\rm hk} = - (f/\rho) \hat{\vec{v}}_\phi$, where $\hat{v}_\phi$ is the versor associated with $\phi$. 
As expected, $\vec{v}_{\rm a}$ is a divergence-free field and indicates that the housekeeping dissipation originates from cyclic trajectories running across the entire circle in a preferential direction, i.e., dragging the particle in the direction of increasing $\phi$, therefore breaking parity symmetry (see Supplementary Information  for a detailed derivation). Yet, the non-equilibrium driving does not affect the steady-state distribution with respect to equilibrium, the system does not explore excited states either stably or through fluctuations, and the Onsager excess dissipation functional $\mathcal{G}_{\ex}$ is minimized to zero. 
In this example, parity symmetry breaking is explicitly linked to the geometry of non-conservative driving, enabling an intuitive identification of adiabatic dissipation that tends to be maximized in a NESS, following $\mathcal{G}_{\rm na}$.

\subsubsection{The Brownian gyrator}\label{sec:gyrator}

As a second paradigmatic example, we study a $2$-dimensional Brownian gyrator, i.e., a diffusive particle in a confining quadratic potential coupled to two reservoirs at different temperatures, $T_x$ and $T_y$, each  acting along one direction (see Fig.~\ref{fig:gyrator}a) \cite{EXARTIER_t}. The Langevin equations describing the system are:
\begin{align}
    \dot{x} &= - \partial_{x} U(x,y)+ \sqrt{2 T_x} \xi_x(t) \;, \nonumber \\
    \dot{y} &= - \partial_{y} U(x,y) +\sqrt{2 T_y} \xi_y(t) \;,
\end{align}
with $\xi$ a Gaussian white noise with zero mean and unit variance, and $T_y = T_x (1+\delta)$. Following Sec.~\ref{sec:baths}, we isolate the interaction part of the potential, $U_{\rm int}$, from confining contributions, $U_x$ and $U_y$:
\begin{gather}
    U = U_{\rm int} + U_x + U_y\;, \nonumber \\
    U_{\rm int} = u\, x y \;, \qquad U_\alpha=\frac{\alpha^2}{2} \;, \quad \alpha = x, y
\end{gather}
where $u$ quantifies the asymmetry of the elliptic potential and  $|u|<1$ to ensure stability. As long as $T_x \neq T_y$, i.e., $\delta \neq 0$, since the mobilities along $x$ and $y$ are the same, the fluctuation-dissipation relation cannot be satisfied and the system breaks detailed balance \cite{kubo1966fluctuation}. The stationary distribution can be explicitly obtained in the following exact form \cite{cerasoli2018asymmetry}:
\begin{eqnarray}
P_{\rm st}&=&\frac{e^{-\FF}}{Z} \;, \quad
\FF=\beta_{\rm eff}\left(\frac{\gamma_{1}}{2}x^2+\frac{\gamma_{2}}{2}y^2+ u \gamma_{3} x y \right)\;, \nonumber
\end{eqnarray}
where $Z$ is the normalization factor, $\beta_{\rm eff}=(T\eta)^{-1}$ an effective inverse temperature, and
\begin{alignat*}{2}
\eta &= 1 + \delta + \frac{u^2}{4} \delta^2 \;, \quad & \gamma_1 &= 2 - \frac{u^2}{2} \delta \;, \\
\gamma_2 &= 1 + \frac{u^2}{2} \delta \;, \quad & \gamma_3 &= 1 + \frac{\delta}{2} \;.
\end{alignat*}
The adiabatic velocity is a solenoidal field vanishing in the availability minimum,  making apparent the parity symmetry breaking. (see Fig.~\ref{fig:gyrator}b). The adiabatic vorticity $w_{\rm st} = \partial_x (v_{\rm a})_y - \partial_y (v_{\rm a})_x$, i.e., the $2D$-curl of the velocity, is connected to the tendency of performing closed trajectories and it reads (see Supplementary Information):
\begin{equation}
    w_{\rm st} = \frac{4 T_x u}{4 T_x T_y + (T_x - T_y)^2 u^2} \left( T_x^2 - T_y^2 \right)\;.
\end{equation}
Following the decomposition in Eq.~\eqref{eq:dec_multiple_bath}, we have:
\begin{gather}
    v^{i}_{\hk} = -\partial_i U_{\rm int} \;, \qquad
    v^i_{\ex, \A} = \nabla^i \FF_{\ex, \A} \;,
\end{gather}
with the excess availability:
\begin{gather}
    \FF_{\ex, \A} = \beta_{\rm eff} \gamma_3 \left(\left(\gamma_{1}-2\right) \frac{x^2}{2} + \left( \gamma_{2}-1 \right) \frac{y^2}{2 \left( 1+\delta \right)} + u x y \right) . \nonumber
\end{gather}
This system fulfills the local equilibrium condition: both housekeeping and excess velocities are divergence-free (harmonic), vanish at the availability minimum, and flow in opposite directions (see Fig.~\ref{fig:gyrator}c). As a result, excited states cannot be stably occupied, although transient explorations occur during fluctuations. The deviation from equilibrium availability arises because the excess velocity induces compensating expansions and contractions along different directions, even though the total divergence is zero. The sum of $\vec{v}_\hk$ and $\vec{v}_{\ex,\A}$ defines the adiabatic velocity, $\vec{v}_\A$, which is a solenoidal field: its circulation does not stem from a non-conservative housekeeping force, as in the previous example, but from the interplay between potential concavity and anisotropic temperatures.
Since this model can be solved analytically at all times, we are able to show in Fig.~\ref{fig:gyrator}d)  the temporal evolution of $\dot{S}_{\rm tot}$, $\dot{S}_{\A}$, and the housekeeping flux $\langle \vec{v} * \vec{v}_{\hk} \rangle$, highlighting that the steady state coincides with a minimization of the total entropy production compatible with the presence of parity symmetry breaking, with the dissipative contribution from excess availability vanishing in the NESS (see Eq.~\eqref{eq:Gfunc}). Furthermore, in Fig.~\ref{fig:gyrator}e, we show that the excess functional $\mathcal{G}_{\ex}$ converges to $(1/2)\dot{S}_{\ex}$ at stationarity. Notice that this value coincides with the minimum over $\vec{v}$ at fixed probability distribution, but $\mathcal{G}_{\ex}$ does not necessarily decrease towards a NESS.

\section{Exploring far from equilibrium with a small noise expansion}\label{sec:noise_exp}
\subsection{Weak-noise limit}
Following the pioneering works of Graham and T\'el \cite{graham1984existence,graham1984weak,graham1985weak}, together with recent developments in Macroscopic Fluctuation Theory \cite{MFT,Garrido_quasi_pot, falasco2023macroscopic}, we propose here that there exist a single noise amplitude $\sigma$ that controls all the elements of the diffusion matrix, so that in the weak-noise regime, the stationary probability distribution exhibits the following limiting behavior:
\begin{eqnarray}\label{eq:noise_large_dev}
  P_{\st} \asymp e^{-\FF^{(0)}/\sigma},
\end{eqnarray}
where $\FF^{(0)}$ is the rescaled zeroth order availability, that constitute also the large deviation or rate function of the system. Furthermore, in this context the availability $\mathcal{F}_{\rm a}$ is a quasi-potential that can be rigorously defined and estimated via path-integral methods \cite{MFT,falasco2023macroscopic}.
From Eq.~\eqref{eq:noise_large_dev}, the adiabatic velocity is then composed by a contribution from the zeroth order availability, i.e., $\vec{v}_{\rm a}=\vec{F} + \nabla \FF^{(0)}$, where, with abuse of notation, we identify $\nabla^i = D^{ij} \partial_j$, without the factor $\sigma$ that rescales the diffusion matrix.
By taking advantage of this general form for the stationary distribution, we uncover the geometrical properties of the velocity $\vec{v}_{\A}$ in the limit of weak noise, i.e., at the leading order in $\sigma$. 
Plugging Eq.~(\ref{eq:noise_large_dev}) into the Fokker-Planck equation, expanding up to the zeroth order in $\sigma$, and equating terms of the same order, we get two consistency equations at stationarity (see appendix \ref{sec:app_noise} for details):
\begin{equation}
 \vec{v}_{\rm a}*\nabla \FF^{(0)}= 0 \;, \qquad \nabla*\vec{v}_{\rm a} = 0 \;.
    \label{topolAdd}
\end{equation}
The first relation states that the $v_{\rm a}$ must be tangent to the quasi-potential contour lines, here identified with $\mathcal{F}^{(0)}$, while the second one dictates that it also has to be a divergence-free field. These conditions imply that, in the minimum of the availability, the current is zero. Following the classification in Sec.~\ref{sec:NESS}, we conclude that, any non-equilibrium system (with additive noise) is locally closed in the weak noise limit, as also remarked by the last line of Table \ref{ness_ch}.
Yet, by employing the house-keeping/excess decomposition,
\begin{eqnarray}
    \vec{v}_{\rm a}&=&\vec{v}_{\hk}+\vec{v}^{(0)}_{\ex,\rm a} \;,\\
    \vec{v}^{(0)}_{\ex,\A}&=&\nabla \FF^{(0)}_{\ex,\A}=\nabla\FF^{(0)}-\beta \nabla U \;,
\end{eqnarray}
we see that the weak noise regime is more general than the locally balanced and locally equilibrium scenarios, since there are no constraints on each component of the velocity.

Notice that, since the form of $P_\st$ in Eq.~\eqref{eq:noise_large_dev} holds without correction in $\sigma$ at equilibrium, i.e., with $\mathcal{F}^{(0)}_{\rm a}/\sigma = \beta U$, every system exhibiting a weak-noise-like stationary distribution independently of the value of $\sigma$ is in local equilibrium, according to the classification in Table \ref{ness_ch}. Indeed, on top of being locally closed, in these systems the extreme points of $\mathcal{F}$ and $U$ have to coincide for any $\sigma$.

\subsection{Beyond weak noise: an additional gradient term in the velocity}\label{sec:gbeyond_weak}
When pushing the system beyond the weak-noise regime, the properties of the velocities are expected to change. This scenario is particularly relevant for mesoscopic thermodynamic systems in which $\sigma$ can be identified with the temperature of the bath. Hence, starting from the rate function in Eq.~\eqref{eq:noise_large_dev}, we consider an additional term in the expansion of the availability:
\begin{eqnarray}\label{eq:expansion_first}
    \mathcal{F} \approx \frac{1}{\sigma}\FF^{(0)}+\FF^{(1)} \;.
\end{eqnarray}
From the decomposition of the velocity in Eq.~\eqref{eq:force_dec}, in addition to the zeroth order term, we obtain a first order contribution in the form of a gradient (see Appendix \ref{sec:app_noise} for details):
\begin{eqnarray}
    \vec{v}_{\A}&=&\vec{v}^{(0)}_{\A}+\sigma \vec{v}^{(1)}_{\A}\\
    \vec{v}^{(0)}_{\A}&=&\vec{F}+\nabla \FF^{(0)}=\vec{v}_{\hk}+\vec{v}_{\ex,\A}^{(0)}\\
    \vec{v}^{(1)}_{\A}&=&\nabla \FF^{(1)}= \nabla \FF-\nabla \FF^{(0)}/\sigma=\vec{v}_{\ex,\A}^{(1)}.
    \label{eq:vel_ex_1}
\end{eqnarray}
Given that $\FF^{(0)}$ is the quasi-potential or large-deviation function of the system, $\FF^{(1)}$ can be identified as the excess availability that vanishes in the weak-noise limit. Therefore, the velocity stemming from its gradient can be identified as an additional excess contribution. 
By solving perturbatively for the stationary solution of Fokker-Planck equation and equating the terms of order $\sigma^{-1}$, we discover that the zeroth order velocity is still orthogonal to the gradient of the quasi-potential:
\begin{eqnarray}\label{eq:first_order-1}
    \vec{v}^{(0)}_{\A}*\nabla\mathcal{F}^{(0)}=0 \;.
\end{eqnarray}
From the terms at following order, $\sigma^0$, we see that its divergence is not zero:
\begin{eqnarray}\label{eq:first_order0}
\nabla*\vec{v}^{(0)}_{\A}=\vec{v}^{(0)}_{\A}*\nabla\FF^{(1)}+\vec{v}^{(1)}_{\A}*\nabla\FF^{(0)} \;.
\end{eqnarray}
Finally, the conditions emerging at order $\sigma$ determine that the first order velocity has a positive divergence:
\begin{eqnarray}\label{eq:first_order1}
\nabla*\vec{v}^{(1)}_{\A}=\vec{v}^{(1)}_{\A}*\nabla\FF^{(1)}=\vec{v}^{(1)}_{\A}*\vec{v}^{(1)}_{\A}\geq 0 \;,
\end{eqnarray}
thus implying that this term will generate repellors. 
\subsection{Geometric and thermodynamics properties beyond weak noise: the effect of the reservoir}\label{sec:geom_beyond_weak}
At stationarity, the time-derivative of the average of the quasi-potential part of the availability has to vanish,
\begin{eqnarray}
\D_t\average{\FF^{(0)}}_{\st}&=&\average{\vec{v}_{\A}*\nabla\FF^{(0)}}_{\st}=0 \;,
\end{eqnarray}
therefore
\begin{eqnarray}
\langle\vec{v}^{(0)}_{\A}
*\nabla\FF^{(0)}\rangle_{\st}&=&- \sigma\average{\vec{v}^{(1)}_{\A}*\nabla\FF^{(0)}}_{\st}
\end{eqnarray}
by using Eq.~\eqref{eq:first_order-1}, we get a new orthogonality condition:
\begin{eqnarray}
    \average{\vec{v}^{(1)}_{\A}*\nabla\FF^{(0)}}_{\st}=0 \;.
\end{eqnarray}
This relation implies that the first order velocity, which consists of only an excess term, is on average orthogonal to the contour lines of the quasi-potential. Similarly, by looking at the excess availability $\FF^{(1)}$, we find that its average change along the zeroth order velocity balances the first order velocity divergence. In fact,
\begin{eqnarray}\label{eq:balance_F1}
\D_t\average{\FF^{(1)}}_{st}&=&\average{\vec{v}_{\A}*\nabla\FF^{(1)}}_{\st}=0
\end{eqnarray}
implies that
\begin{equation}\label{eq:balance_2}
\begin{split}
\average{\vec{v}^{(0)}_{\A}*\nabla\FF^{(1)}}_{\st} &= - \sigma \average{\vec{v}^{(1)}_{\A}*\vec{v}^{(1)}_{\A}}_{\st} \\
&=- \sigma\average{\nabla*\vec{v}^{(1)}_{\A}}_{\st} 
\;,
\end{split}
\end{equation}
where we used the fact that $\vec{v}^{(1)}_\A = \vec{\nabla} \FF^{(1)}$ and Eq.~\eqref{eq:first_order1} for the last equality. Notice that, by using Eq.~\eqref{eq:balance_F1} and Eq.~\eqref{eq:first_order0}, the relation appearing in Eq.~\eqref{eq:balance_2} is equivalent to:
\begin{eqnarray}\label{eq:div_balance}
\average{\nabla*\vec{v}^{(0)}_{\A}}_{\st} = -\sigma\average{\nabla*\vec{v}^{(1)}_{\A}}_{\st}\leq 0 \;,
\end{eqnarray}
that can be obtained by directly averaging the divergence of the total velocity, and remembering that $\average{\nabla*\vec{v}_{\A}}_{\st}=0$ at stationarity. These conditions, taken together, imply that the average fluxes of $\vec{v}^{(0)}_\A$ and $\vec{v}^{(1)}_\A$ need to match, with a consequent non-zero flux of the first-order availability that shifts the extreme points of the system. This flux, shown in Eq.~\eqref{eq:balance_2}, can be interpreted as emerging from local couplings with the reservoir that have been ignored in the weak-noise regime since any system in this limiting scenario results to be locally closed.
These geometric properties have an impact on the thermodynamics of the system and, in particular, they affect the excess-housekeeping decomposition of the entropy production. As a first observation, let us remark that, from Eqs.~\eqref{eq:stationary} and \eqref{eq:fp_div}, one has that the stationary probability can now locally expand and contract, i.e., the flow is compressible, and hence the adiabatic availability con increase or decrease along the velocity lines. As a result, single trajectories can now travel between different availability valleys. In second place, we notice that the divergence of the excess velocity gets modified by the coupling with the reservoir encoded in $\FF^{(1)}$:
\begin{eqnarray}\label{eq:div_vex}
\nabla*\vec{v}_{\ex,\A}= \nabla*\vec{v}^{(0)}_{\ex,\A}+\sigma \vec{v}^{(1)}_{\ex,\A}*\vec{v}^{(1)}_{\ex,\A}.
\end{eqnarray}
while the housekeeping contribution stays unchanged because of Eq.~\eqref{eq:vel_ex_1}. This substantiates the fact that, beyond the weak-noise limit, the system is no longer locally closed since Eq.~\eqref{eq:closness} cannot be satisfied at any order in $\sigma$. This result highlights the crucial role of the noise amplitude in enabling the identification of a coupling with the reservoir through the geometric properties of the system, i.e., from locally closed to open by increasing $\sigma$.

\begin{figure*}[t]
    \centering
    \includegraphics[width=2\columnwidth]{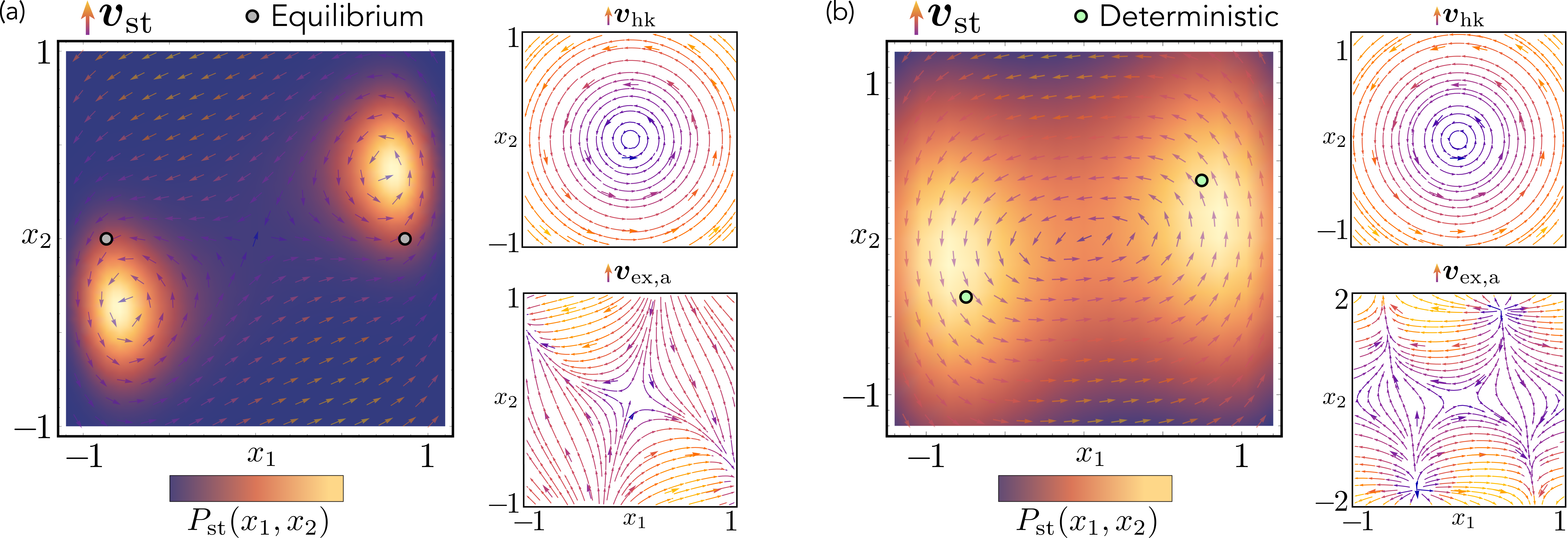}
    \caption{\textbf{Beyond weak noise for a Brownian particle in bistable potential and shear flow.} 
    (a) Contour plot of the steady-state distribution in the weak-noise limit with the adiabatic velocity as a vector field in color-code (brighter arrow corresponds to stronger field). The minima of the equilibrium potential are shown: they do not coincide with the peaks of the availability, leading to a non-zero excess availability. We report both the corresponding excess velocity as a stream plot in color-code and the housekeeping component that coincides with the shear flow. (b) Same as in panel (a), but for the strong-noise scenario. Notice that the  availability minima do not coincide with the deterministic solution, and the excess velocity shows the presence of repellors at $(x_1, x_2) \approx \pm (0.5, 1.8)$. 
    However, the housekeeping component is unchanged, since it only depends on external non-equilibrium conditions. Parameters are: $a = 1, b = 1.5, \gamma = 1$, and $T = 0.1$ (weak noise) or $1$ (strong noise).}
    \label{fig:shearflow}
\end{figure*}

Another effect of going beyond the weak-noise regime is that the excess entropy acquires new terms:
\begin{eqnarray}
    \dot{S}_{\ex,\A}= \dot{S}^{(0)}_{\ex,\rm a}+\dot{S}^{(1)}_{\ex,\rm a}+2 \average{\vec{v}_{\ex,\rm a}^{(0)}*\vec{v}_{\ex,\rm a}^{(1)}} \;,
\end{eqnarray}
where we introduced
\begin{eqnarray}
    \dot{S}^{(0)}_{\ex,\rm a}&=& \frac{1}{\sigma}\average{\vec{v}_{\ex,\rm a}^{(0)}*\vec{v}_{\ex,\rm a}^{(0)}},\quad\dot{S}^{(1)}_{\ex,\rm a}=  \sigma\average{\vec{v}_{\ex,\rm a}^{(1)}*\vec{v}_{\ex,\rm a}^{(1)}} \;. \nonumber
\end{eqnarray}
By using Eq.~\eqref{eq:first_order1}, it is immediate to realize that the first order excess entropy production measures the flow of its corresponding velocity:
\begin{eqnarray}\label{eq:entropy_prod1_div}
   \dot{S}^{(1),\st}_{\ex,\rm a}=\sigma\average{\nabla*\vec{v}^{(1)}_{\rm ex,\A}}_{\st}\geq 0
\end{eqnarray}
Eq.(\ref{eq:entropy_prod1_div}) expresses the fact that the interactions with the reservoir create local dissipative probability expansions that tend to increase the availability.
From Eq.~\eqref{eq:div_balance}, expressing the divergence of $\vec{v}^{(1)}$ from Eq.~\eqref{eq:balance_2}, and then using the excess/housekeeping decomposition for the zeroth-order velocity, we obtain:
\begin{eqnarray}\label{eq:availability_balnce_1}
\average{\nabla^2\FF^{(0)}_{\ex,\A}}_{\st}&=&-\average{\nabla*\vec{v}_{\hk}^{(0)}}_{\st}- S^{(1),\st}_{\ex,\A}
\end{eqnarray}
This equation specializes Eq.~\eqref{eq:possion-non-eq} for open systems at the first order in $\sigma$. It clearly shows that the deviation of availability concavity from equilibrium are dictated not only by the housekeeping velocity, reflecting external non-equilibrium constrains, but also by an additional excess entropy production term stemming from the system-reservoir thermodynamic coupling.

As a last point, from the definition of the excess entropy production, after simple manipulation, we get:
\begin{equation}
    \dot{S}_{\ex,\rm a}^{\st} = \dot{S}_{\ex,\rm a}^{(0),\st}-
 \dot{S}^{(1),\st}_{\ex,\rm a}-2 \average{ v_{\hk}*\nabla\FF^{(1)}}_{\st}\geq 0 \;.
\end{equation}
Hence, depending on the magnitude of the house-keeping flux of $\FF^{(1)}$, the magnitude of the excess dissipation can change, while always satisfying the global constraints in Eqs.~\eqref{eq:excess_house_keeping_bal} and \eqref{eq:excess_house_keeping_limits}.

\subsection{Example: Brownian particle in a bistable potential and shear flow}
As an example, we consider a Brownian particle immersed in a bath at temperature $T$ under the effect of a double well potential $U(\vec{x})=a x^4_1-b x^2_1+x^2_2$ and a shear flow $\vec{f}=\gamma\{-x_2,x_1\}$ (by components):
\begin{eqnarray}
    \dot{x}_i=-\partial_iU+f_i+\sqrt{2T}\xi_i \;.
\end{eqnarray}
In the absence of flow ($\gamma=0$), the system reach the equilibrium distribution $P_{\eq}\sim {\rm exp}(-U/T)$.
Conversely, in the presence of driving a large deviation function in the form of Eq.~\eqref{eq:noise_large_dev} exists and can ve numerically evaluated, as shown in~\cite{santolin2024bridgingfreidlinwentzelllargedeviations}. Even if $\FF^{(0)}$ is not accessible analytically, we study the deterministic dynamics through a fixed points analysis, and perform numeric integration of the Fokker-Planck equation. In in the strong driving limit, i.e., $\gamma^2> 4b$, the availability will have a single minimum in $\vec{x}=0$, while by lowering $\gamma$, $\vec{x} = 0$ becomes a maximum and two minima appear:
\begin{eqnarray}\label{eq:deterministic}
    x^*_1=\pm  \frac{1}{2}\sqrt{\frac{4\tilde{b}-\gamma^2}{2a}} \;, \qquad x^*_2=\gamma \frac{x^*_1}{2} \;,
\end{eqnarray}
where $\tilde{b}=b - 6 a {\rm Var}(x_1)$ indicates how the parameters are modified by the effect of noise through the variance of $x_1$ (the correction is of order $T$ for small temperatures, see \cite{santolin2024bridgingfreidlinwentzelllargedeviations} for details). By studying the system in the weak-noise limit, we see that the non-equilibrium conditions, i.e., the presence of the shear flow, modifies the availability landscape and the minima of $\FF$ become equal to those in the deterministic limit, Eq.~\eqref{eq:deterministic} \cite{wang_rev}. They do not coincide with those of the potential $U$, so that the system is not in local equilibrium conditions (see Fig.~\ref{fig:shearflow}a). As showed in Sec.~\eqref{sec:non_cons_force}, $\vec{v}_{\hk}$ coincides with the shear flow $\vec{f}$, i.e., it is divergence-less, and, as a consequence, the system falls in the locally balanced scenario. In Fig.~\ref{fig:shearflow}a, we show the stationary distribution and the housekeeping/excess velocities in the weak-noise regime.
However, when we increase the temperature that controls the noise amplitude, new phenomena appear. The system is now locally open and exchanges energy with the bath. As such, $\vec{v}_{ex,\rm a}$ is not necessarily harmonic, repellors appear in the system, and the availability minima do not coincide with the deterministic solution  (see Fig.~\eqref{fig:shearflow}b). However, the divergence-less nature of the house-keeping velocity ensures that the two excess contributions balance each other, i.e., while the system is locally open, it remains globally closed:
\begin{equation}
\average{\nabla*\vec{v}^{(0)}_{\ex,a}}_{\st} = - \dot{S}^{(1),\st}_{\ex,\A} = -\sigma\average{\nabla*\vec{v}^{(1)}_{\ex,\A}} \leq 0
\end{equation}
or, equivalently,
\begin{equation}
\average{\nabla^2\FF^{(0)}_{\ex,\A}}_{\st} = - \dot{S}^{(1),\st}_{\ex,\A} = -\sigma\average{\nabla^2\FF^{(1)}_{ex,\A}}_{\st} \leq 0 \;,
\end{equation}
where the excess entropy production measures the energy flow from the bath to the system. These equalities have been derived by using Eqs.~\eqref{eq:div_vex} and \eqref{eq:availability_balnce_1}. Therefore, an increase in temperature generates dissipative expansions of probability and the system is doing work at the single trajectory level by populating excited states. In summary, the non-trivial interplay between temperature bath, equilibrium potential, and energy flow has the effect of rising the system's typical energies and increasing the excess dissipation.

\section{Multiplicative fluctuations}\label{sec:multiplicative}
\subsection{A new velocity stemming from curved geometry}
The proposed framework naturally extends beyond the case of additive noise to the challenging scenario of multiplicative fluctuations, where a formulation of both Onsager and Glansdorff-Prigogine principles is lacking, as far as we know. In general, state-dependent diffusion coefficients may emerge in a variety of systems, from Brownian motion in inhomogenous media or constrained on curved surfaces \cite{VANKAMPEN1988673, van1986brownian, Landauer1988MotionOO} to chemical reaction networks affected by thermal gradients \cite{busiello2021dissipation, dass2021equilibrium, liang2022emergent} and fluctuating environments \cite{nicoletti2021mutual}. Moreover, finite-size fluctuations are inherently multiplicative and important in fluctuating hydrodynamics \cite{Garrido_2021}, field theories \cite{te2020classical,cornalba2019regularized} and population dynamics \cite{azaele2006dynamical,prr_evo_sireci}.

We start from the following general Langevin equation:
\begin{equation}
    \dot{\vec{x}} = \vec{F}(x) + G(x) \vec{\xi}(t)
    \label{LangevinMult} \;,
\end{equation}
with $\vec{\xi}$ a vector of Gaussian white noises with zero mean and correlation matrix $2\sigma \mathbb{1}$ (the identity matrix), where $\sigma$ controls noise amplitude. Here, $G$ is a state-dependent matrix, hence the total diffusion matrix is $D=\sigma G^{T}G$. For $G = \mathbb{1}$, this system reduces to the additive noise case. This new scenario extends and generalizes our approach by considering again the inverse of the diffusion matrix to be the metric of our space, which can now be curved because of the space-dependence of $D$ (that we assume to be invertible). 
Following the classic works by Graham \cite{Graham1977CovariantFO,GRAHAM_itp}, and a more recent deep analysis by Polettini \cite{polettini2013generally}, we use standard tools of differential geometry to recast the Fokker-Planck equation in a fully covariant form (i.e., coordinate independent), see Appendix \ref{sec:geom_mult} for details. First, we need to upgrade the standard derivatives to covariant ones:
\begin{eqnarray}
    \nabla_{j} u^{i}=\partial_{j}u^{i}+\Gamma^{i}_{j k}u^{k},
\end{eqnarray}
where $\Gamma$ is the Levi-Civita connection, i.e. the tensor that encodes the effect of the curvature on the derivative, while the first contribution measures the transport in the tangent space (that in a flat manifold coincides with the usual configuration space). Recall that the covariant derivative of a scalar field coincides with the partial derivative.
The contravariant derivative, or gradient, is obtained from the covariant derivative: $\nabla^{i}v^{k}=D^{ij}\nabla_{j}v^{k}$. As a consequence, the divergence now depends on the diffusion matrix determinant $\vert D\vert$:
\begin{eqnarray}\label{eq:cov_div}
    \nabla*\vec{v}=\sqrt{\vert D \vert}\partial_{i}\left(\frac{v^{i}}{\sqrt{\vert D\vert}}\right)= \partial_iv^i-\Gamma_iv^i\;.
\end{eqnarray}
with $\Gamma_i=\Gamma^{k}_{ik}=-\partial_i\log\sqrt{\vert D\vert}$.
By defining the invariant probability Q and the invariant measure $d\Omega$, i.e., $Q=P\sqrt{\vert D\vert}$, and $d\Omega = dx/\sqrt{\vert D\vert}$, one can rewrite the Fokker-Planck equation in a fully covariant form, interpreting the Langevin equation with the Stratonovich ($\alpha=0$) or Ito ($\alpha=1$) prescription:
\begin{equation}\label{eq:FPE_cov}
    \partial_t Q=-\nabla_{i}J^{i},\quad J^{i}= \tilde{F}^{i}Q-\sigma \nabla^{i}Q \;.
\end{equation}
In this expression, the force is modified by a term that we call diffusive velocity, $\vec{v}_{D}$:
\begin{equation}
\label{eq:mod_force}
    \tilde{F}^{i} = F^{i}+ v_{\rm D}^i \;, \quad
   v_{\rm D}^i= D^{ij}\omega_j-\alpha\partial_{j}G^{i}_{l}G^{j}_{k}\delta^{lk} \;,
\end{equation}
that is composed by a vector field $\vec{\omega}$ that measures the breaking of detailed balance condition on the diffusive force $G$:
\begin{equation}\label{eq:omega}
    \omega_{j} = G^l_k\left(\partial_jG^k_l-\partial_l G^k_j\right) \;,
\end{equation}
and by the Ito phoretic term, $G^{T}\partial G$. The vector field $\vec{\omega}$ stems directly from the configuration space geometry: it vanishes when $G$ can be written as the derivative of a collection of vector fields, $G^i_j=-\partial_j\psi^i$, that corresponds to the condition of zero curvature of the metric induced by the diffusion matrix (see Ref.~\cite{Graham1977CovariantFO} and Appendix \ref{sec:appendix_exp_mult}).
This modification of the force field directly implies also a change in the velocity decomposition:
\begin{equation}
    \vec{v}_{\A} = \vec{\tilde{F}}-\nabla Q= \vec{F}-\nabla Q+\vec{v}_{D}.
\end{equation}
where the diffusive (geometry-driven) velocity emerges.
\subsection{Geometric and thermodynamics properties of multiplicative fluctuations}
In the case of multiplicative fluctuations, equilibrium cannot generally be reached, making the identification of a potential energy less clear. Nevertheless, the zero current condition implies the following detailed balance relationship for the modified force in Eq.~\eqref{eq:mod_force}:
\begin{eqnarray}\label{eq:db2}
    \partial_i\tilde{F}_j=\partial_j\tilde{F}_{i} \;.
\end{eqnarray}
It is then trivial to see that, once again, any gradient field does not break this condition. Therefore, it can be collected into the excess velocity and interpreted as a contribution to a pseudo-equilibrium distribution, i.e., a modification of the internal organization of system's states. Yet, the diffusive velocity has a complex structure that makes challenging the general identification of the excess and housekeeping terms, along with their geometric properties. This is due mainly to the presence of $\vec{\omega}$ that introduces cross dependencies into the fluctuations. 
Let us investigate two different scenarios. In the first one, the diffusion matrix is space-dependent, but the diffusive forces satisfy the potential condition $G^i_{k}=-\partial_{k}\psi^{i}$.
As a result, $\vec{\omega}$ vanishes, the geometry is flat and it is always possible to find a set of coordinates $\vec{x}'=G^{-1}\vec{x}$ (called \emph{harmonic}), where the system has additive noise~\cite{Graham1977CovariantFO, polettini2013generally}.
Focusing on the Ito prescription ($\alpha=1$), in this simple setting  the diffusive velocity turns out to be a gradient
\begin{eqnarray}
v_{D}^i=\frac{1}{2}\partial^{i}\left(\partial_j\psi_{k}\partial_{z}\psi_{l}\delta^{kl}\delta^{jz}\right)=v_{\ex,\rm D}^i \;.
\end{eqnarray}
Hence, it can be absorbed into the excess part of the velocity and does not contribute to the breaking of detailed balance, Eq.~\eqref{eq:db2}. An intuitive understanding comes from considering a system subjected to a thermal gradient along each direction, $T_i(x_i)$ \cite{VANKAMPEN1988673,celani2012anomalous}. If the gradients cause independent fluctuations on the system, the diffusion matrix is diagonal $D = G^{T}G = {\rm diag}(T_i(x_i))$ and $G_{ii} = \partial_i \psi(x_i)$. Therefore, the velocity is a ``phoretic'' term:
\begin{eqnarray}\label{eq:phoretic_v_ito}
    v_{D}^i=-\nabla^i\log\sqrt{\vert D\vert} =-\delta^{ik}\partial_k T_i\left(x_{i}\right)0=\Gamma^i \;.
\end{eqnarray}
stemming directly from the connection $\Gamma$ and contributing to the covariant divergence Eq.~\eqref{eq:cov_div}.
By considering the noise expansion for the adiabatic availability, we obtain a covariant version of the equations in Sec.\ref{sec:gbeyond_weak} with this phoretic term entering in the first order velocity $\vec{\tilde{v}}^{(1)}_\A = \nabla\FF^{(1)}+\vec{v}_{\rm D}$ (see Appendix \ref{sec:appendix_exp_mult}). As a result, the 
first order excess dissipation is given by the velocity divergence in tangent space:
\begin{equation}
\begin{split}
\dot{\tilde{S}}^{(1)}_{\ex,\A} &=\sigma\average{\vec{\tilde{v}}^{(1)}_{\rm a}*\vec{\tilde{v}}^{(1)}_{\rm a}}_{\rm st}=\sigma\average{\partial*\vec{\tilde{v}}^{(1)}_{\A}}_{\st} \geq 0\;.
\end{split}
\end{equation}
We can conclude that the system is open and the phoretic velocity will drive the system to stably occupy ``excited'' states.
It is important to remark that, in this case, even though the curvature vanishes, the connection is not zero and directly generates the diffusive velocity. Therefore, $\vec{v}_{\rm D}$ plays the role of a fictitious force, arising solely from the use of a coordinate system in which the metric (and thus the diffusion tensor) varies in space. While in Newtonian frames fictitious forces originate from accelerated or curvilinear coordinates, here they emerge from spatially varying diffusion, with the effective drift reflecting how the local diffusive motion changes across space. Let us briefly discuss the more general case in which $\omega\neq 0$. This is generally expected when the coordinates exhibit coupled diffusivity or when the diffusion tensor is non-diagonal, encompassing scenarios such as Brownian particles in magnetic fields \cite{brownian_magnetic}, diffusion on curved surfaces \cite{van1986brownian}, and interacting active particle systems \cite{cates_review, mackay2025emergentdynamicsactivesystems}.
In this scenario, $\vec{v}_{D}$ arises intrinsically from the curvature, and cannot be eliminated by a change of coordinates. This is analogous to a static gravitational field likewise producing thermal gradients according to the Ehrenfest-Tolman effect \cite{Rovelli_2011}. The drift in this $\omega \neq 0$ case contributes simultaneously to housekeeping and excess velocities, since the distinction between gradient and rotational components becomes entangled with curvature. A precise separation is nontrivial, and we leave this interesting but challenging problem for future work.

\subsection{Non-adiabatic functional in the presence of multiplicative fluctuations}

The non-adiabatic non-equilibrium functional can be derived also in the presence of multiplicative fluctuations, i.e., for a system evolving according to Eq.~\eqref{LangevinMult}. By computing the Kullback-Leibler divergence between $P(x,t)$ and the NESS, we obtain:
\begin{equation}
\begin{split}
     \mathcal{G}_{\na} &= \average{ \vec{v} *\vec{v} } - \average{\vec{v} * \vec{v}^{(0)}_{\hk}} -\average{ \vec{v} * \nabla\FF^{(0)}_{\ex,\A} }\\
     &\qquad\qquad {}-\sigma\average{ \vec{v} * \nabla\FF^{(1)}_{\ex,\A} }
    -\sigma \average{ \vec{v} * \vec{v}_{D}} \;.
    \label{eq:GfuncMult}
    \end{split}
\end{equation}
In analogy to Eq.~\eqref{eq:Gfunc}, the first contribution quantifies the total entropy production, which tends to be minimized as the system evolves toward stationarity. The second term amounts to the dissipated heat due to (deterministic) non-equilibrium conditions. In the case $\vec{\omega}=0$, the last three terms constitute the excess dissipation, and quantify, in order of appearance, the distance from equilibrium of the quasi-potential part of the availabiltiy, the effect of the bath in modifying the landscape, and finally, the effect of multiplicative fluctuations. This last contribution, as we have seen, is proportional to the derivative of $|D|$ and vanishes in the limit of additive noise. In particular, assuming the validity of the Einstein relationship for every point in space, it can be readily shown that this last term is $\propto  \nabla T$, resembling an additional dissipation arising from thermophoretic effects and due to the necessity of transporting heat \cite{liang2022emergent,celani2012anomalous}. Geometrically, this term signals a breaking of translational invariance induced by state-dependent diffusion, which forces the system into excited states. In the more general case $\vec{\omega}\neq 0$, the contribution of $\vec{v}_{\rm D}$ will modify also the housekeeping dissipation, inducing a further non-conservation of angular momentum and breaking of parity symmetry. Despite similar dissipative terms have been already obtained in the presence of a temperature gradient \cite{celani2012anomalous}, our derivation is much more general and applies even outside the realm of purely thermodynamic systems. 
The Onsager excess dissipation functional remains valid with multiplicative noise, and its minimization under the orthogonality constraint implies that the system not only relaxes toward equilibrium but also adapts to the curvature of configuration space. Thus, the housekeeping/excess splitting naturally reflects distinct broken symmetries in trajectory space, extending the Onsager and Glansdorff–Prigogine principles  up to the first order in the noise amplitude and in the presence of multiplicative noise, an intricate and often analytically inaccessible scenario.

\subsection{Example: Underdamped particle  in a\\temperature gradient}\label{sec:underdamped}
In this section, we extend our approach to the case of an underdamped system in which the multiplicative noise originates from a temperature gradient. For simplicity, we consider the following $1$-dimensional underdamped dynamics of a Brownian particle in the position-velocity phase space $(x,\V)$ in the presence of a temperature gradient $T(x)$:
\begin{eqnarray}
\dot{x} &=& \V \;,\\
\dot{\V} &=& - \frac{\gamma}{m} \V - \frac{1}{m} \partial_x U + f + \frac{1}{m} \sqrt{2 \gamma T(x)} \xi(t)\;,
\end{eqnarray}
where $\gamma$ represents the friction, $m$ the mass, $f$ a non-conservative force even under time reversal, the potential $U = u x^2$ for simplicity, and the diffusion coefficient satisfies the Einstein's relation $D(x) = \gamma T(x)$ at each point in space. We can define the timescale of the friction $\tau = m/\gamma$ so that the overdamped limit is attained for $\tau \to 0$, when velocities equilibrate infinitely fast.

The current velocity is now a $2$-dimensional vector in the phase space and can be split into a reversible $\vec{v}_{\rm rev}$ and an irreversible component $\vec{v}_{\rm irr}$, according to the behavior they exhibit under time reversal operation \cite{dechant2018entropic}: 
\begin{equation}
\begin{split}
    \vec{v}_{\rm rev} &= \left\{ \V, - \frac{1}{m} \partial_x U(x) + \frac{f}{m} \right\}, \\
    \vec{v}_{\rm irr} &= \Big\{ 0, - \frac{1}{\tau} \V - \frac{1}{m \tau} T(x) \partial_\V \underbrace{\log(P(x,\V,t))}_{\FF} \Big\} \;. 
    \end{split}
\end{equation}
At equilibrium, when $T(x) \equiv T$ and $f=0$, the system cannot support any dissipative flux. Solving the Fokker-Planck equation associated with the Langevin dynamics without a temperature gradient, we obtain the equilibrium distribution
\begin{equation}
    P_{\eq}(x,\V) \sim  e^{-\frac{m \V^2}{2 T} - \frac{U(x)}{2 T}} = e^{-\beta E} \;,
\end{equation}
where $\beta = 1/k_B T$ with $k_B = 1$, and $E$ is the total energy of the system, as expected.

\begin{figure}
    \centering
    \includegraphics[width=\columnwidth]{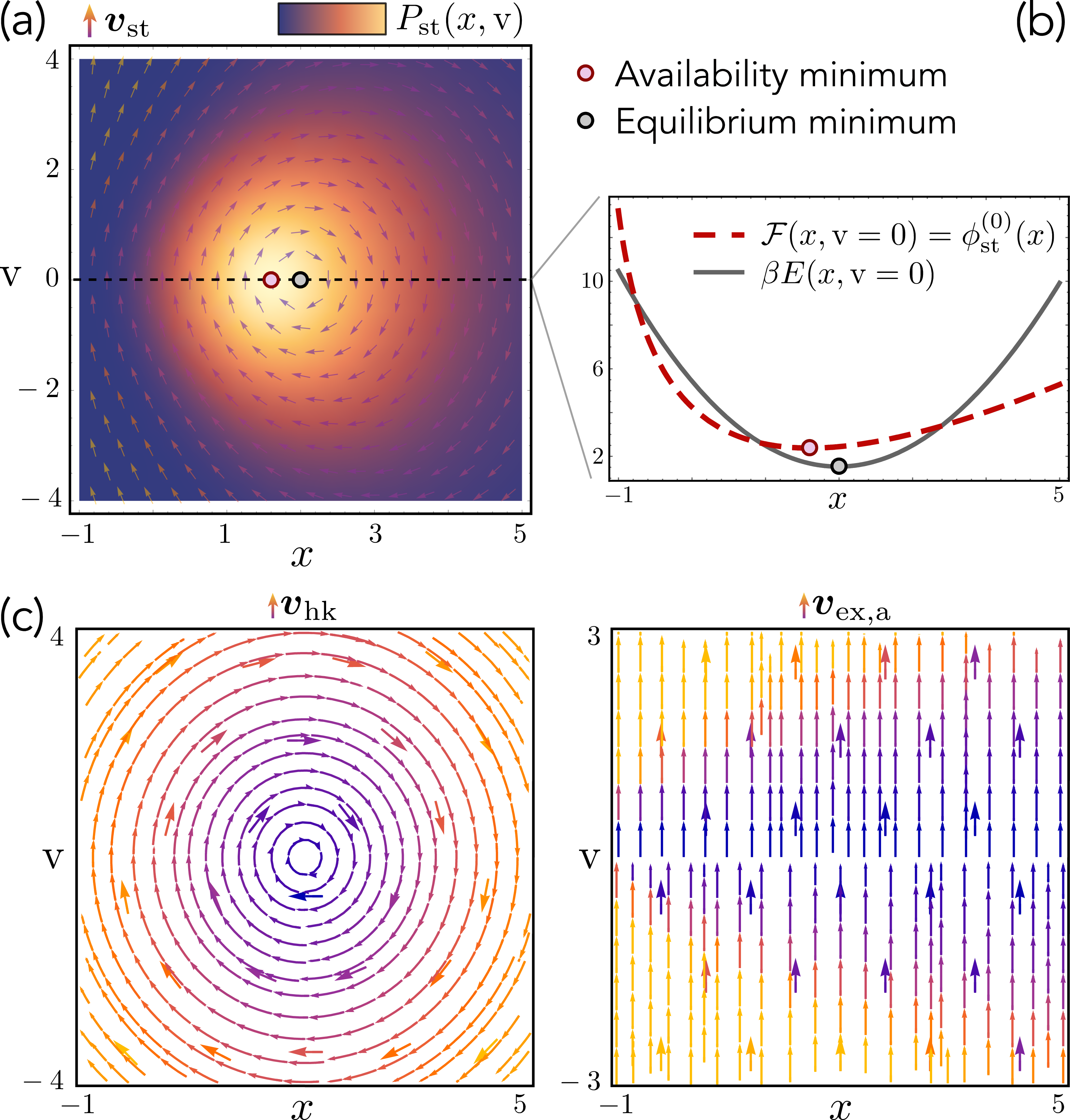}
    \caption{\textbf{Geometry of velocities for underdamped dynamics with thermal gradient.} (a) The stationary solution of the underdamped Fokker-Planck equation up to the first order in $\tau$ is shown (Eq.~\eqref{eq:P_under}), for a linear thermal gradient $T(x) = T_0(1 + \alpha x)$ and a quadratic potential $U = u(x - \mu)^2$. Arrows represent the stationary velocity in color-code (brighter arrows indicate stronger field). (b) Comparison between availability and equilibrium distribution for ${\rm v} = 0$. Notice that the availability along this line coincides with $\phi^{(0)}_{\rm st}(x)$ (Eq.~\eqref{eq:P_under}). The availability minimum does not coincide with the equilibrium one. (c) Streamplot in color-code of housekeeping (reversible) and excess (irreversible) velocities. In this figure, the spatial domain is $x \in [-1/\alpha, +\inf]$ and we set all parameters so that the probability distribution and its derivative vanish at the boundaries. The model parameters are: $m = 1$, $\gamma = 0.01$, $\mu = u = 2$, $T_0 = 1$, $\alpha = 0.8$.}
    \label{fig3}
\end{figure}

In analogy to the two bath scenario discussed in Section \ref{sec:baths}, we define the excess velocity by fixing $P_{\eq}$ as reference distribution for every point in space, i.e., $P_\eq|_{\beta \to \beta(x)}$. The resulting decomposition is:
\begin{equation}
    \vec{v}_\hk = \vec{v}_{\rm rev} \qquad \vec{v}_{\ex} = \nabla (\FF - \beta(x) E) = \vec{v}_{\rm irr}
\end{equation}
A crucial observation to obtain this correspondence is that only $\V$ has a metric term given by the diffusion, as the system is effectively $1$-dimensional. Therefore, the gradient only acts on the $\V$-space. As said above, this decomposition is arbitrary, but it is informed (and substantiated) by the physical decomposition of the velocity into reversible and irreversible components. As a consequence, the excess part accounts for the deviation of the availability from the equilibrium condition. It is non-zero and, as such, it allows the system to stably explore excited stated due to the presence of a thermal gradient. Conversely, the housekeeping velocity is divergence-free, contains the non-conservative force pushing the system out of equilibrium, and also reflects the presence of fluxes in the full phase space that vanish in the overdamped limit. Indeed, $\V$ acts as a nonequilibrium driving, since collapsing the system into the $x$-space (i.e., fast relaxation of $\V$ leading to the overdamped regime) will result into an effective equilibrium distribution.
When the system is pushed out of equilibrium by a temperature gradient, the Fokker-Planck equation can be solved at any time by expanding in powers of $\tau$, i.e., close to the overdamped regime. Up to the first order, the time-dependent non-equilibrium solution reads (see Supplementary Information for the derivation):
\begin{equation}
\begin{split}
    &P^{(\tau)}(x,\V,t) = e^{-\frac{m \V^2}{2T}} \bigg\{ \Phi^{(0)} - \tau \bigg[ \V \partial_x \Phi^{(0)} +\\
    &\qquad +\bigg( \frac{m \V^3}{6 T^2} 
     + ~\frac{\V}{T} \bigg) \Phi^{(0)} \partial_x T + \frac{\V}{T} \Phi^{(0)} \partial_x U \bigg] \bigg\},
    \label{eq:P_under}
    \end{split}
\end{equation}
where $\Phi^{(0)}(x,t) = \phi^{(0)}(x,t) \sqrt{2 \pi T(x)/m}$, with $\phi^{(0)}$ satisfying the zero-flux overdamped Fokker-Planck equation with Ito prescription (see Fig.~\ref{fig3}a-b), as dictated by the time-scale separation employed \cite{liang2022emergent,kupferman2004ito}. First, we notice that the form of the solution in Eq.~\eqref{eq:P_under} goes beyond the weak noise regime, thereby showing a peak that is different from the equilibrium one (see Fig.~\eqref{fig3}b).
In Fig.~\ref{fig3}c, we show the components of the stationary velocity stemming from this solution.

Despite an apparent similarity, considering the entire phase space leads also to crucial differences in the thermodynamic meaning of our results. We start by computing the non-adiabatic functional in this case:
\begin{eqnarray}\label{eq:func_under}
    \mathcal{G}_{\rm na} &=& -\frac{d D_{KL}}{dt}= \langle \vec{v} * \left( \vec{v} - \vec{v}_{\rm a} \right) \rangle \\
    &=&\langle \vec{v}_{\rm irr} * \left( \vec{v}_{\rm irr} - \vec{v}_{\rm irr, a} \right) \rangle + 
 ~\langle \vec{v}_{\rm rev} * \left( \vec{v}_{\rm irr} - \vec{v}_{\rm irr, a} \right) \rangle \;. \nonumber
\end{eqnarray}
Notice that the irreversible current velocity is the only term contributing to the steady-state entropy production rate of an underdamped system, i.e., $\dot{S}_{\rm tot} = \langle \vec{v}_{\rm irr} * \vec{v}_{\rm irr} \rangle$ \cite{celani2012anomalous,dechant2018entropic}. This is the first difference with respect to the previous overdamped scenario. In fact, in this system, we observe a purely excess dissipation, since $\vec{v}_{\rm irr}$ coincides with $\vec{v}_{\ex}$. Analogously, we can provide a physical interpretation for all other terms appearing in the functional. We have that $\langle \vec{v}_{\rm irr} * \vec{v}_{\rm irr,st}\rangle = \langle \vec{v}_{\rm irr} * \vec{v}_{\rm \ex,\A}\rangle$, the irreversible part of the flux of the adiabatic excess availability, can be interpreted as the average heat dissipation rate (divided by the temperature) into the environment at steady state $\langle \dot{Q}_{\rm st}/T(x) \rangle$. Finally, the novel term stemming from the angle between the two velocities $\langle \vec{v}_{\rm rev} * \vec{v}_{\rm irr}\rangle = \langle \vec{v}_{\rm hk} * \vec{v}_{\rm \ex}\rangle$, i.e., the reversible part of the excess availability flux, also has a physical interpretation, as it quantifies the average work rate performed on the system divided by the temperature $\langle \dot{W}/T(x) \rangle$ (see Supplementary Information). Putting all terms together, the non-equilibrium function can be written as:
\begin{equation}
    \mathcal{G}_{\na} = \bigg\langle \frac{\dot{Q}+\dot{W}}{T(x)} - \frac{\dot{Q}_{\rm a}+\dot{W}_{\rm a}}{T(x)} \bigg\rangle = \bigg\langle \frac{\dot{E} - \dot{E}_{\rm a}}{T(x)} \bigg\rangle.
\end{equation}
Hence, the non-equilibrium functional for underdamped systems corresponds to the average rate of excess energy change, rescaled by the temperature. 
This result sheds new light on the intrinsic physical meaning of the non-equilibrium functional when the full phase space is considered. An in-depth analysis of this intricate scenario is left for future studies.

\section{Generalization to many-body systems: geometry of free-energy transduction in coupled molecular oscillators}\label{sec:osc}
\begin{figure*}[t]
    \centering
    \includegraphics[width=\textwidth]{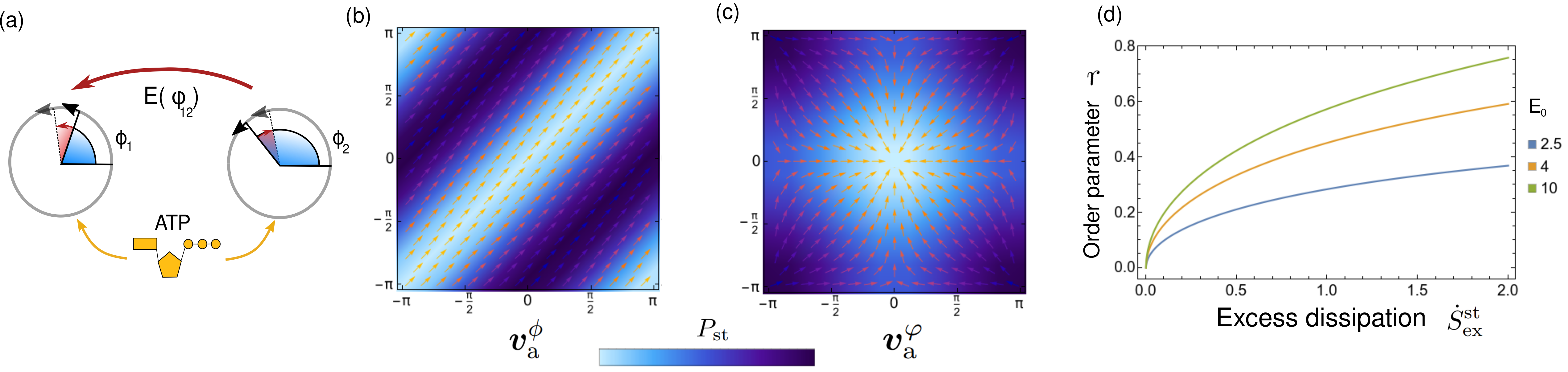}
    \caption{\textbf{The geometry of current velocities for coupled molecular oscillators.} 
    (a) Pictorial sketch of two coupled molecular biochemical oscillators. Each oscillator is represented by a phase $\phi_i\in [0,2\pi]$ (in shaded blue color) that grows with a constant velocity $ke_g$ (solid black arrows) fueled by ATP consumption. Furthermore, oscillators can interact with a phase exchange potential $E_{12}(\varphi_{12})$ depending on their phase difference $\varphi_{12}=\phi_1-\phi_2$ (in shaded red), eventually leading to anti-correlated phase changes (small red arrows).
    (b) Stream plot of $v^{\phi}_{\A}$  for a typical oscillator with phase $\phi_1$ as a function of the collective phase $\psi$  with the heat map of $P_{\st}(\theta_1)$, $\theta_1=\phi_1-\psi$ in the mean-field limit in the synchronized phase. Here, $v^{\phi}_{\A}$ drives the synchronized oscillations (yellow arrows and light-blue regions, high probability density) but also indicates that interactions tend to destabilize them (dark blue regions of extremely low probability). 
    (c) Stream plot of $v^{\varphi}_{\A}$ for two typical oscillators with the heat map of $P_{\st}(\theta_{1},\theta_{2})$, where , with $\psi$ the mean-field collective phase. $v^{\varphi}_{\st}$ tends to reduce the oscillators phase differences $\varphi$, stabilizing the dynamics and promoting synchronization. 
    (d) Plot of the synchronization order parameter as a function of the excess entropy for different values of $E_0$. Model parameters are $k=0.5, e_g=4\pi, E_0=8, \Omega=1$.}
    \label{fig:velocity_osc}
\end{figure*}
Our framework is able to provide physical insights in even more general settings. To present a concrete example, we now employ it to analyze a many-body system, i.e., a model of $N$ coupled biochemical oscillators introduced in \cite{Zhang2019TheEC}.
Consider $N$ oscillators, each one with a phase $\phi_i \in [0,2\pi]$ and undergoing a stochastic dynamics that advances its phase. The resulting phase procession rate is equal to $k e_g$, where $e_g$ is a typical thermodynamic force driving the system out of equilibrium, while the fluctuation strength is proportional to $k$, a typical microscopic reaction rate. The bath temperature has been put to $1$ for simplicity.  Then, we include the possibility that two oscillators can exchange a phase through the interaction potential, $E_{ij}=E(\phi_i-\phi_j)=-E_0\cos\left(\phi_i-\phi_j\right)$, with $E_0$ the coupling energy (see Fig.\ref{fig:velocity_osc} (a) for a pictorial illustration).
This system can hence be described by $N$ coupled Langevin equations:
\begin{equation}\label{eq:lang_phase}
    \dot{\phi}_i = k e_g+\eta_{i}-\frac{\Omega}{N}\sum_{i<j}(E'_{ij}+\xi_{ij}(t))
\end{equation}
where $\Omega$ is a typical frequency. Here $\eta_i$ and $\xi_{ij}$ are two different zero-mean Gaussian noises respectively stemming from single-oscillator dynamics and generated by interactions between oscillators, with variances:
\begin{eqnarray}
   \langle \eta_{i}(t)\eta_{j}(t')\rangle&=&2k\delta_{ij}\delta(t-t') \nonumber \\
   \langle \xi_{ij}(t)\xi_{kl}(t')\rangle&=&\frac{2\Omega}{N}[\delta_{ik}\delta_{jl}-\delta_{il}\delta_{jk}]\delta(t-t') \nonumber \;.
\end{eqnarray}
For the detailed derivation of this model from a microscopic description, along with its biological motivation, we refer to \cite{Zhang2019TheEC}. Let us emphasize that this example goes beyond the setting presented in the first section because of the presence of two different currents, and hence two noise parameters, and we do not have a natural reference equilibrium distribution. Nevertheless, we show that our framework is still valid and indeed it illuminates the system phase transition. One can calculate the exact stationary solution of this system and identify the adiabatic availability as:
\begin{eqnarray}
  P_{\st}=\frac{e^{-\beta_{\rm eff}\FF}}{Z},\quad \FF=N^{-1}\sum_{i<j}E^{ij}
\end{eqnarray}
with $\beta_{\rm eff}=1+k/\Omega$ (see also the Supplementary Information for its derivation) and the total energy plays the role of the availability. Given the many-body nature of the problem, a very particular thermodynamics emerges. First, we notice that Eq.~\eqref{eq:lang_phase} is equivalent to $N$ equations describing phase changes, $\phi_i$, coupled to $N^2$ equations capturing phase differences, $\varphi_{ij} = \phi_i - \phi_j$. With this splitting, the whole system can be described using the formalism introduced in Section \ref{sec:geom_dis}, but in two different spaces. First, we identify housekeeping and excess velocity associated with the single oscillator phase change:
 \begin{eqnarray}\label{eq:vel_oscil}
\vec{v}^{\phi}_{\A}&=&\vec{v}^{\phi}_{\hk}+\vec{v}^{\phi}_{\ex,\A},\quad
        \vec{v}^{\phi}_{\hk}=ke_g,\nonumber\\ (\vec{v}^{\phi}_{\ex})^{i}&=&\nabla^i_{\phi}\FF= \frac{k\beta_{\rm eff}}{N}\partial_{j}E^{ij}
\end{eqnarray}
where the gradient   is defined as $\nabla^i_{\phi}=k\delta^{ij}\partial_j$  and scalar product is indicated as $a\cdot b=k^{-1}a_{i}b^{i}$. Second, in the space of the phase difference, indicating the gradient as $ \nabla^{ij}_{\varphi}=(\Omega/N)\left(\partial^i-\partial^j\right)$, we obtain the velocities:
\begin{eqnarray}
        &&\vec{v}^{\varphi}_{\A}= \vec{v}^{\varphi}_{\hk}+
        \vec{v}^{\varphi}_{\ex, \A}, \quad  (v^{\varphi}_{\hk})^{ij}=-\frac{\Omega}{N}E'^{ij}\nonumber\\
      &&(v^{\varphi}_{\ex,\A})^{ij}=
         \nabla^{ij}_{\varphi}\mathcal{F}=
    \frac{\Omega\beta_{\rm eff}}{N^2}\partial_{k}\left(E^{ik}-E^{jk}\right).
\end{eqnarray}
and the scalar product is $a*b=(\Omega/N)^{-1}\sum_{i<j}a^{ij}b^{ij}$.
The geometrical picture is very rich and generalizes our previous results. In each space, there is a housekeeping part that represents the injected energy, i.e., the ATP for single phases, where $\vec{v}_\hk$ is harmonic and $\nabla_{\phi} \cdot \vec{v}^{\phi}_{\hk}=0$, and the interaction potential for phase differences, where $\nabla_{\varphi} * \vec{v}^{\varphi}_{\hk}\neq 0$. The excess parts are given by the derivatives of the availabilities in each space, with the one associated with $\varphi$ representing the energy difference between oscillators. Finally, both adiabatic velocities have non-zero divergences, and their values balance:
\begin{eqnarray}\label{eq:div_osc}
 \nabla_{\phi}\cdot \vec{v}^{\phi}_{\A}=-\nabla_{\varphi}* \vec{v}^{\varphi}_{\A}=k\beta_{\rm eff}\nabla_{\phi}^2\FF_{\A}. 
\end{eqnarray}
By performing a mean-field limit, valid for an infinite number of oscillators ($N \to\infty$), one discovers that the system can undergo a transition from an asynchronous to a coherent phase by varying the coupling frequency $\Omega$ (see the original work \cite{Zhang2019TheEC} for more details). A plot of the system velocities for two mean-field oscillators in the synchronized phase is presented in Fig.~\ref{fig:velocity_osc} (see the Supplementary Information for the explicit expressions). It shows intuitively that, while $\vec{v}^{\phi}$ drives the phase of each single oscillator to complete a cycle (non-zero vorticity but not divergence free), $\vec{v}^{\varphi}$ tends to reduce the phase differences to zero and hence to synchronize the system (i.e., a multiple dipole structure emerges akin to a group of charges in an electric field, zero-vorticity). In such a limit we can also compute the velocities average divergence 
\begin{eqnarray}\label{eq:av_div_osc}
 \average{\nabla*\vec{v}^{\varphi}_{\A}}_{\st}&=&k\beta_{\rm eff}E_0 r=k\beta_{\rm eff}\average{\nabla^2_{\phi}\FF_{\A}}_{\rm st}\geq 0,\;
 \end{eqnarray}
 and the various  entropy production terms:
 \begin{eqnarray}
     \dot{S}^{\st}_{\hk} &=& ke_g^2+\frac{\Omega E_0}{\beta_{\rm eff}}\left(1-\frac{1}{\beta_{\rm eff} E_0} \right), \nonumber\\
    \dot{S}^{\st}_{\ex} &=& \Omega E_0r^2= -\FF^{\rm flux}_{\hk}\geq 0 \;,
    \label{eq:mean_field_entropy}
\end{eqnarray}
which contribute to the total dissipation, as for Eqs.~\eqref{eq:excess_house_keeping_bal}. Eqs.~\eqref{eq:av_div_osc} and \eqref{eq:mean_field_entropy} give a complete thermodynamic characterization of the synchronization phase transition: the order parameter $r\in [0,1]$ stems from the velocity divergence (i.e. the availability Laplacian), so that the housekeeping dissipation correctly quantifies energy inputs (depending only on the control parameters), as anticipated above, 
while the excess one quantifies exactly the resulting emergent behavior, as it originates from the transduction work done by the oscillators and, as such, is proportional to the square of the order parameter.
Following the Prigogine and Onsager principles, we can deduce that, when approaching the NESS, the system minimizes the total dissipation compatibly with aligning the phase velocity of each oscillator with both inputs from ATP and all other oscillators, i.e., $\dot{S}^{\phi}_{\A}=\langle \vec{v}^{\phi}\cdot \vec{v}^{\phi}_{\A}\rangle$, and the coherence between them, i.e., $\dot{S}^{\varphi}_{\A}=\langle \vec{v}^{\varphi}* \vec{v}^{\varphi}_{\A}\rangle$.
This suggests that, in many-body systems, the non-trivial currents geometry, i.e., the divergence of $\vec{v}$ and excess dissipation reflect the transduction of free energy from one components to the others, similarly to the view developed by T. Hill \cite{Hill1989FreeEnergyTransduction, wachtel2022free}.

\section{Conclusions and perspectives}
\subsection{Summary}
How order emerges far from thermodynamic equilibrium remains a far-reaching unsolved problem. In this manuscript, we shed light on how geometrical and symmetry-breaking properties of currents in non-equilibrium steady states (NESSs) emerge from the violation of detailed balance. We propose a framework capable of isolating the conditions responsible for breaking detailed balance (the housekeeping part) and the system’s internal response to them (the excess), both at the level of current velocities and thermodynamic dissipation. From this decomposition, we derive several results, including an upper bound for the stationary entropy production and a geometric characterization of NESSs. In particular, we show that the housekeeping component is related to the tendency to perform cycles in trajectory space (breaking parity symmetry), while the excess velocity captures the conservative (gradient) part of the dynamics, pushing the system towards selected ``excited'' regions of the state space and, as such, reflecting the appearance of internal organization. The excess entropy production quantifies the distance from equilibrium of the system’s availability,  playing the role of a non-equilibrium functional that should be minimized to derive the correct NESS velocity geometry. Consequently, any NESS can be understood as the least dissipative state compatible with maximal dissipation along these structures in trajectory space. By performing a weak-noise expansion of the system availability, we obtain a general and model-free characterization of velocity current geometry, along with its thermodynamic consequences, highlighting how environmental stochasticity increases the excess contribution. Furthermore, extending these results to systems with multiplicative fluctuations—--a largely uncharted area—--reveals that more complex scenarios can emerge. We uncover the structure of both excess and non-adiabatic dissipation in this case, showing how the system breaks both parity and translational symmetry in trajectory space due to the presence of an additional phoretic drift. Finally, applying our ideas to many-body systems with coupled molecular oscillations reveals the geometric properties of free-energy transduction and non-equilibrium phase transitions when multiple currents are simultaneously at play.
\subsection{Towards a thermodynamics of selection and symmetry breaking}
\begin{figure}[t]
    \centering
  \includegraphics[width=\linewidth]{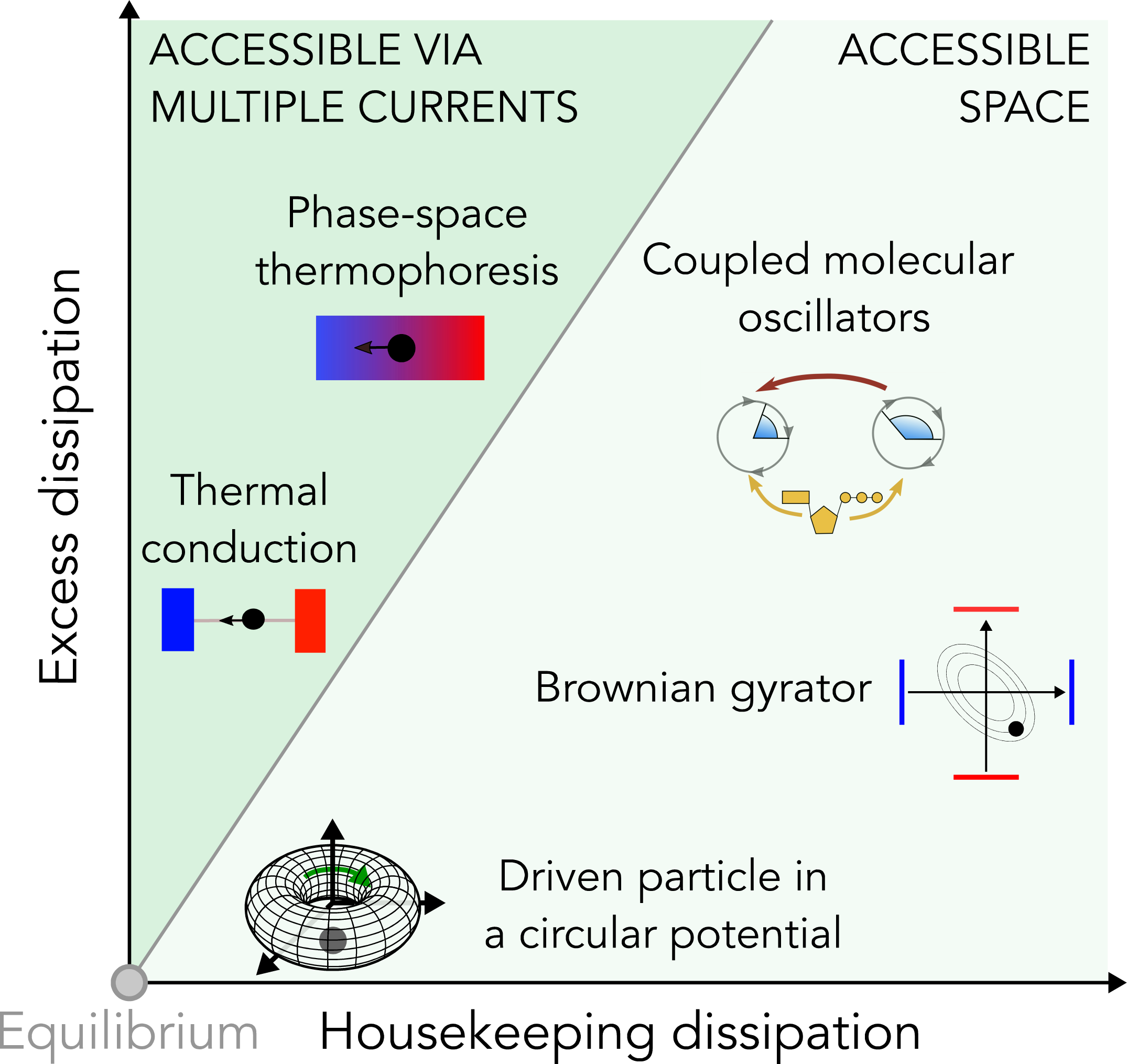}
    \caption{\textbf{NESS classification based on housekeeping and excess dissipation}. We report the examples studied along this manuscript in the diagram, with their position qualitatively representing the thermodynamic classification on the base of Table~\ref{ness_ch}. Local equilibrium system such as the driven particle in a circular potential or the Brownian gyrator have zero or modest excess dissipation, and the velocity geometry is dominated by the housekeeping contribution breaking parity. They all have to live in the accessible space where $\dot{S}_{\hk} \geq \dot{S}_{\ex,\A}$, since they are characterized by a single physical current. For all these systems, the gray line and the gray dot indicate equilibrium conditions. However, open systems like a particle connected to many heat baths or an underdamped particle in a temperature gradient sustain the emergence of multiple currents that might enable the access to the whole space, even exhibiting a purely excess dissipation associated with a translational symmetry breaking (see Sec.~\ref{sec:underdamped}). Finally, a many-body open system like a biochemical oscillator exhibits an increasing excess dissipation when approaching a synchronization phase transition.}
    \label{fig:classification}
\end{figure}
An intuitive understanding of our result might come from interpreting the onset of these geometrical structures as a mechanism to cope with the external constraints that keep the system out of equilibrium. Indeed, the system selectively dissipates energy through these ``large-scale'' complex structures, which are also responsible for the breaking of parity and translational symmetry (in the most general case). Following inspiration by Zia \cite{Zia_2006}, the excess and housekeeping dissipation can be used to geometrically classify NESSs, as recapitulated in Table 1 and in Fig.\ref{fig:classification}. In this work, we verified our results mainly in cases where it is possible to access the analytical form of the stationary probability distribution, at least in an approximate sense. To advance further, we plan to investigate the problem numerically, using new methodologies capable of reducing the error in velocity estimation. On the analytical side, the properties of velocity fields could eventually be studied using renormalization group techniques \cite{RG_tu1,RG_tu2}, large-deviation theory \cite{cycle_random_walk}, or by focusing on topological properties. Finally, on a more conceptual level, our approach calls for a deeper understanding of non-equilibrium thermodynamics through the lens of gauge transformations, geometry and information theory \cite{Polettini_2012, wang_gauge, ito2022information, nicoletti2021mutual}.
Beyond these fundamental questions, our framework promises shows great promise as a versatile tool for addressing different interdisciplinary challenges, as already shown by flux/landscape approaches \cite{wang_rev}.
Active matter represents a promising field of application for the presented framework. In the simplest case of an active Brownian particle \cite{dabelow2019irreversibility}, for example, a solution can be readily obtained at all times and closely resembles that of the $2D$ Brownian gyrator. The conceptual leap, in this context, is that emergent dissipative cycles arise in the space describing both the particle and the bath, making physical interpretation more challenging \cite{active_engine,obyrne}. Future work may further explore this direction, together with the role of non-reciprocal interactions \cite{fruchart2021non}.
More generally, following the many-body example presented here, our framework could pave the way for constructing a thermodynamics of non-equilibrium phase transitions, where individual components break symmetry due to energy injection but coordinate to spontaneously induce macroscopic order \cite{yu2022energy,falasco2023macroscopic}. Along the same lines, the idea that a form of selection can naturally occur out of equilibrium due to dissipation-driven processes is gaining momentum in the fields of physical and biological/metabolic chemistry \cite{busiello2021dissipation,dass2021equilibrium, Liu2023LightdrivenED, blokhuis_evo, goyal2023closed}, suggesting a possible conceptual bridge between thermodynamics and evolutionary dynamics \cite{PRICE, prr_evo_sireci, rao_evo, dill_selection}.  
To further explore this analogy, we leave for future investigation the extension of the framework to discrete-state dynamics (e.g., master equations), in line with existing exploratory efforts in this direction \cite{ito2022information, ito_non_linear, dalcengio, RG_tu1, RG_tu2, liang_symmetry}.
In summary, our work generalizes the principle of excess entropy production minimization while providing a physical and geometric interpretation to it. Owing to its generality, it paves the way for future theoretical investigations and offers a unified approach to understanding selection phenomena across different contexts as a consequence of symmetry-breaking processes driven by non-equilibrium dissipation.

\acknowledgments
M.S. acknowledges G. Jona-Lasinio for early inspiration in studying symmetry breaking in non-equilibrium statistical physics and M. Muñoz for initiating to the quasi-potential formalism. D.M.B. thanks M. Ciarchi (Retoño) for illuminating discussions on symmetry breaking.
We are extremely grateful to M.A. Muñoz, E. Roldan, A. Maritan, P.I. Hurtado, P. Garrido, C. Maes and A. Goyal for important discussions and insights. M.S. has been supported  Grant No. PID2023-149174NB-I00 financed by the Spanish Ministry and Agencia Estatal de Investigación MICIU/AEI/10.13039/501100011033 and EDRF  funds, by Daniel R. Amor (ÉNS-PSL) through the grant ANR-22-CPJ2-0064-01 and by the Human Frontier Science Program (HFSP) Postdoctoral Fellowship LT0022/2025-L.  D.M.B. is funded by the program STARS@UNIPD with the project ``ActiveInfo''.
\appendix 
\section{Weak noise expansion and covariant formulation  
of the Fokker-Planck equation}
\subsection{Perturbative solution of the Fokker-Planck equation in additve noise}\label{sec:app_noise}
If we assume a weak noise ansatz for the stationary probability:
\begin{eqnarray}
    P_{\st}\asymp e^{-\FF^{(0)}/\sigma},
\end{eqnarray}
the statioanary condition reads:
\begin{eqnarray}\label{eq:stationary_app1}
\partial_{t}P_{\rm st}&=&
\sigma^{-1}P_{\rm st}\left(F^{i}+D^{ij}\partial_{j}\FF^{(0)}\right)\partial_{i} \FF^{(0)}+\nonumber\\&-&\left(\partial_{i}F^{i}+D^{ij}\partial_{i}\partial_{j}\FF^{(0)}\right)P_{\rm st} =0.
\end{eqnarray}
Following  various approaches \cite{graham1984weak,MFT,Garrido_quasi_pot}, we perturbatively solve Eq.~\eqref{eq:stationary_app1} by equating to zero each order in $\sigma$. The first contribution is of order $1/\sigma$ and reads:
\begin{eqnarray}\label{eq:order0_app1}
(F^{i}+D^{ij}\partial_{j}\FF^{(0)})\partial_{i}\FF^{(0)}=0 \;,
\end{eqnarray}
while, the second (and last) contribution is of order $\sigma^0$:
\begin{eqnarray}\label{eq:order1_app1}
\partial_{i}F^{i}+D^{ij}\partial_{i}\partial_{j}\FF^{(0)}=0 \;.
\end{eqnarray}
By using the definition of gradient, divergence and adiabatic velocity,
\begin{eqnarray}
    \vec{v}_{\rm a}=\vec{F}+\nabla\FF^{(0)} \;.
\end{eqnarray}
Equations \eqref{eq:order0_app1} and \eqref{eq:order1_app1} in vector form read:
\begin{eqnarray}
   \vec{v}_{\rm a}*\nabla\FF^{(0)}=0,\qquad \nabla*\vec{v}_{\rm a}=0 \;.
\end{eqnarray}
If we consider an additional order in the availability expansion, we have
\begin{eqnarray}\label{eq:noise_ansatz_appendix}
    P_{\st}\approx e^{-\FF^{(0)}/\sigma-\FF^{(1)}} \;,
\end{eqnarray}
so that the velocity can be split into two parts:
\begin{eqnarray}
    \vec{v}^{(0)}_{\rm a}=\vec{F}+\nabla\FF^{(0)}, \quad    \vec{v}^{(1)}_{\rm a}=\nabla\FF^{(1)} \;.
\end{eqnarray}
The stationary condition reads:
\begin{eqnarray}\label{eq:stationary_app2}
\partial_{t}P_{\rm st}&=&
\sigma^{-1}P_{\rm st}\left(F^{i}+D^{ij}\partial_{j}\FF^{(0)}\right)\partial_{i} \FF^{(0)}\\
&+&\Big(\partial_{i}\left(F^{i}+D^{ij}\partial_{j}\FF^{(0)}\right)+\nonumber\\
&-&(F^{i}+D^{ij}\partial_{j}\FF^{(0)})\partial_{i}\FF^{(1)}
-D^{ij}\partial_{j}\FF^{(1)}\partial_{i} \FF^{(0)}\Big)P_{\rm st} \nonumber\\
&+&\sigma\left(-D^{ij}\partial_{i}\partial_j\FF^{(1)}+D^{ij}\partial_{i}\FF^{(1)}\partial_{j}\FF^{(1)}\right)P_{\rm st}=0.\nonumber
\end{eqnarray}
The dominant order $\sigma^{-1}$ is not modified with respect to Eq.~\eqref{eq:order0_app1}:
\begin{eqnarray}\label{eq:order_zero_perp1}
   \vec{v}^{(0)}_{\rm a}*\nabla\FF^{(0)}=0.
\end{eqnarray}
On the other hand, the order $\sigma^0$ is modified to:
\begin{eqnarray}\label{eq:first_v}
&-&(F^{i}+D^{ij}\partial_{j}\FF^{(0)})\partial_{i}\FF^{(1)}+\partial_{i}\left(F^{i}+D^{ij}\partial_{j}\FF^{(0)}\right)+\nonumber\\
&-&D^{ij}\partial_{j}\FF^{(1)}\partial_{i} \FF^{(0)}=0 \;.
\end{eqnarray}
In vector forms it reads:
\begin{eqnarray}
    \nabla*\vec{v}^{(0)}_{\rm a}=\vec{v}^{(0)}_{\rm a}*\vec{v}^{(1)}_{\rm a}+\vec{v}^{(1)}_{\rm a}*\nabla\FF^{(0)}
\end{eqnarray}
Finally, the order $\sigma$ reads
\begin{gather}\label{eq:second_v}
-D^{ij}\partial_{i}\partial_j\FF^{(1)}+D^{ij}\partial_{i}\FF^{(1)}\partial_{j}\FF^{(1)}=0\\
\nabla*\vec{v}^{(1)}_{\rm a }=\vec{v}^{(1)}_{\rm a }*\vec{v}^{(1)}_{\rm a} \;.
\end{gather}

\subsection{Covariant formulation of the Fokker-Planck equation with multiplicative noise}\label{sec:geom_mult}
We first consider the covariant formulation of the Fokker-Planck equation in the Ito scheme, and secondly in the Stratonovich case. Our exposition follows various works by Graham \cite{Graham1977CovariantFO,GRAHAM_itp}.

In the Ito case, the Fokker-Planck equation reads:
\begin{eqnarray}
    \partial_t P&=&-\partial_i\left(F^iP-\partial_jD^{ij}P\right)\nonumber\\
    &=&-\partial_i\left(\left(F^i-\partial_{j}D^{ij}\right)P-D^{ij}\partial_jP\right).
\end{eqnarray}
As in Sec.~\ref{sec:geom_dis}, we take the inverse of the diffusion matrix as a metric to introduce the covariant derivative:
\begin{eqnarray}
    \nabla_{j} u^{i}=\partial_{j}u^{i}+\Gamma^{i}_{j k}u^{k} \;,
\end{eqnarray}
where $\Gamma$ is the affine connection. By imposing the parallel transport of the metric,
\begin{eqnarray}
     \nabla_{j} D^{ij}=0 \;,
\end{eqnarray}
the connection is determined to be:
\begin{eqnarray}
    \Gamma^{\alpha}_{\beta \kappa}=\frac{1}{2}D^{\alpha\mu}\left(\partial_{\beta}D_{\mu,\kappa}+\partial_{\kappa}D_{\mu\alpha}-\partial_{\mu}D_{\alpha\beta}\right) \;.
\end{eqnarray}
The covariant divergence is defined as:
\begin{eqnarray}\label{eq:div_cov_app}
   \nabla*\vec{v}&=& \nabla_{i}v^{i}=\sqrt{\vert D\vert}\partial_{i}\left(\frac{v^{i}}{\sqrt{\vert D\vert}}\right)\nonumber\\&=&\partial_i v^i-\Gamma_iv_i \;,
\end{eqnarray}
where $\vert D\vert$ is the diffusion matrix determinant and $\Gamma_i=\Gamma^{k}_{ik}$.
The notation in terms of the diffusive scalar product follows from:
\begin{eqnarray}
\nabla*\vec{v}&=&\nabla^{i}D_{ij}v^{j}=D^{ik}\nabla_{k}D_{ij}v^{j} \nonumber \\
&=&\nabla_jv^{j}+D^{ik}v^j\nabla_kD_{ij}=\nabla_jv^{j}+{D_{i}}^{k}v_j\nabla_kD^{ij} \nonumber\\ &=&\nabla_jv^{j}+v_j\nabla_iD^{ij}=\nabla_jv^{j} \;.
\end{eqnarray}
The Laplace-Beltrami operator, i.e., the divergence of a gradient of scalar function, is defined as:
\begin{eqnarray}\label{eq:laplace_beltrami}
    \nabla*\nabla \phi=\sqrt{\vert D\vert}\partial_{i}\left(\frac{D^{ij}\partial_j\phi}{\sqrt{\vert D\vert}}\right).
\end{eqnarray}
Now, by defining the invariant probability Q and the invariant measure $\D\Omega$
\begin{eqnarray}\label{eq:invariant_measure_app}
    Q=P\sqrt{\vert D\vert},\quad \D\Omega=\frac{\D\vec{x}}{\sqrt{\vert D\vert}} \;,
\end{eqnarray}
one can rewrite the Fokker-Planck equation in a fully covariant form. First, we need to set the position of the diffusion matrix in a compatible way with the Laplace-Beltrami operator. This will produce an inhomogeneity in the force field. In the Ito prescription, this reads:
\begin{gather}\label{eq:dis_ito_der}
    \sqrt{\vert D\vert}\partial_{i}\partial_jD^{ij}P=\sqrt{\vert D\vert}\partial_{i}\partial_j\left(\frac{D^{ij}Q}{\sqrt{\vert D\vert }}\right)\\
    =\sqrt{\vert D\vert}\partial_{i}\left(\frac{D^{ij}\partial_jQ}{\sqrt{\vert D\vert }}\right)+\sqrt{\vert D\vert }\partial_i\left(Q\partial_j\left(\frac{D^{ij}}{\sqrt{\vert D\vert }}\right)\right)\nonumber
\end{gather}
By rewriting now everything using Eqs.~\eqref{eq:div_cov_app}, \eqref{eq:laplace_beltrami}, and \eqref{eq:invariant_measure_app}, we arrive at:
\begin{eqnarray}
    \partial_t Q&=&-\sqrt{\vert D \vert}\partial_i\left(\frac{\tilde{F}^iQ}{\sqrt{\vert D\vert }}-\frac{D^{ij}\partial_jQ}{\sqrt{\vert D\vert }}\right)\nonumber\\
    &=&-\nabla*J \;,
\end{eqnarray}
where
\begin{eqnarray}
    J^i&=&\tilde{F}^{i}Q-D^{ij}\partial_jQ \;, \nonumber\\
   \tilde{F}^{i} &=& F^{i}-\sqrt{\vert D\vert}\partial_{j}\left(\frac{D^{ij}}{\sqrt{\vert D\vert }}\right)=F^i+v^i_{\rm D} \;
\end{eqnarray}
This new term appearing in the force, that we name $\vec{v}_{\rm D}$, is the covariant divergence of the diffusion matrix. By applying the definition of the covariant divergence, Eq.~\eqref{eq:cov_div}, it can be rewritten as:
\begin{eqnarray}
v_{\rm D}^i&=&\partial_j D^{ij}-\Gamma^i\nonumber\\
\Gamma^i&=& \Gamma^k_{jk}D^{ij}=-\nabla^{i}\log\sqrt{\vert D\vert} \;.
\end{eqnarray}
Finally, by using Eq.~\eqref{eq:omega} for the first term and after some manipulations, it can be decomposed as the sum of a first contribution stemming from the vector $\vec{\omega}$, and a second one from the derivative of the diffusive forces $G$:
\begin{eqnarray}\label{eq:v_d_ito}
    v_{\rm D}^i &=& D^{ij}\omega_j-\partial_{j}G^{i}_{l}G^{j}_{k}\delta^{lk} \;.
\end{eqnarray}
On the other hand, in the case of the Stratonovich prescription, we have first to rewrite the Fokker-Planck to make manifest the presence of the diffusive matrix (as we did in the Ito case): 
\begin{eqnarray}\label{eq:fp_strato}
    \partial_t P&=&-\partial_i\left(FP-G^{i}_{b}\partial_j\left(G^{j}_a\delta^{ab}P\right)\right)\nonumber\\
    &=&-\partial_i\left(\left(F^i+\partial_{j}G^{i}_{k}G^{j}_{k}\right)P-\partial_jD^{ij}P\right),
\end{eqnarray}
and then apply Eq.(\ref{eq:dis_ito_der}).It is immediate to see that the additional drift term in the force in Eq.(\ref{eq:fp_strato}) counterbalances the second term in Eq.~\eqref{eq:v_d_ito}, leading to the following expression for the diffusive velocity:
\begin{eqnarray}\label{eq:v_d_strato}
    v_{\rm D}^i=D^{ij}\omega_j.
\end{eqnarray}
The combination of Eqs.~\eqref{eq:v_d_ito} and \eqref{eq:v_d_strato}, with the introduction of a  parameter $\alpha=0$, $1$, gives Eq.~\eqref{eq:mod_force} in the main text. Now, let us comment on the curvature and its relation with the diffusive velocity. As usual, the curvature can be quantified by the Ricci scalar in terms of the affine connection:
\begin{eqnarray}
    R=D^{ij}\left(\partial_l\Gamma^{l}_{ij}-\partial_j\Gamma^{l}_{il}+\Gamma^{k}_{ij}\Gamma^l_{lk}-\Gamma^{k}_{il}\Gamma^l_{jk}\right).
\end{eqnarray}
The diffusive forces $G$ are the metric frames, and intuitively they tell how the variable feel the noise ``kicks'' at a certain configuration \cite{polettini2013generally}. The vector $\omega$ is called anholonomity vector and measures how $G$ behaves when parallel transported along a closed loop, i.e., the anisotropy of fluctuations. This is equivalent to measuring the curvature of the associated manifold. Hence, one defines the anholomity tensor as
\begin{eqnarray}
    \omega^{c}_{ab}=G^i_aG^j_b\left(\partial_i G^c_j-\partial_j G^c_i\right)
\end{eqnarray}
which quantifies how a motion along the directions $a$ and $b$ generates anisotropy in $c$. The correspondent anholomity vector capturing the net twist only along the direction $a$ is thus obtained by contraction. Then, one can rewrite the scalar curvature as:
\begin{eqnarray}
    R= \frac{1}{4} \omega_{abc} \omega^{abc} - \frac{1}{2} \omega_{abc} \omega^{bca} - \omega_a{}^{ab} \omega^c{}_{cb} \;.
\end{eqnarray}\\

\subsection{Perturbative solution of the  covariant Fokker-Planck equation with multiplicative fluctuations}\label{sec:appendix_exp_mult}
We consider a noise expansion for the invariant stationary probability:
\begin{eqnarray}
 Q_{\st}\approx e^{-\FF^{(0)}/\sigma-\FF^{(1)}} \;,
\end{eqnarray}
and solve order by order in sigma as in Sec.~\ref{sec:app_noise}. The results are equivalent to the previous expansions but stated in curved geometry, and with the addition of the ``phoretic'' term in the first order velocity,
\begin{eqnarray}
    \vec{\tilde{v}}^{(1)}_{\rm a}&=&\nabla\FF^{(1)}+\vec{v}_{D} \;.
\end{eqnarray}
The dominant order is still given by Eq.~\eqref{eq:order_zero_perp1},
while the following orders read:
\begin{eqnarray*}
    \nabla*\vec{v}^{(0)}_{\rm a}&=&\vec{v}^{(0)}_{\rm a}*\nabla\FF^{(1)}+\vec{\tilde{v}}^{(1)}*\nabla\FF^{(0)}\\
      \nabla*\vec{\tilde{v}}^{(1)}_{\rm a}&=&\vec{\tilde{v}}^{(1)}_{\rm a}*\nabla\FF^{(1)}=\vec{\tilde{v}}^{(1)}_{\rm a}*\left(\vec{\tilde{v}}^{(1)}_{\rm a}-\vec{v}_{D}\right) \;.
\end{eqnarray*}
By taking the average, the last equation leads to the first order entropy production:
\begin{eqnarray}\label{eq:s1_d_sup}
    \dot{\tilde{S}}^{(1)}&=& \sigma\average{ \vec{\tilde{v}}^{(1)}_{\rm a}*\vec{\tilde{v}}^{(1)}_{\rm a}}\nonumber\\
    &=&\sigma\average{ \nabla*\vec{\tilde{v}}^{(1)}_{\rm a}}+\sigma\average{\vec{v}_{D}*\vec{\tilde{v}}^{(1)}_{\rm a}}\geq 0 \;.
\end{eqnarray}
In the case of vanishing curvature, $\vec{\omega}=0$ and  it is immediate to show that the Ito diffusive velocity Eq.~\eqref{eq:v_d_ito} reduces to the contracted connection (see Eq.~\eqref{eq:phoretic_v_ito}). Hence by applying the definition of covariant divergence, Eq.~\eqref{eq:cov_div}, in Eq.~\eqref{eq:s1_d_sup}, one obtains
\begin{eqnarray}
       \dot{\tilde{S}}^{(1)}_{\rm ex,\rm a}=\sigma\average{ \partial*\vec{\tilde{v}}^{(1)}_{\rm a}} \;,
\end{eqnarray}
coherently with the fact that the manifold is flat.

\end{document}